\documentclass[pra,twocolumn,nofootinbib]{revtex4}

\usepackage{graphicx}
\usepackage{amssymb}
\usepackage{amsmath}
\usepackage{amsfonts}
\usepackage{epstopdf}
\usepackage{epsfig}
\usepackage{wrapfig}

\begin{document}

\title{Internal Consistency of Fault-Tolerant Quantum Error Correction
  in Light of Rigorous Derivations of the Quantum Markovian Limit}
\author{Robert Alicki$^{(1)}$, Daniel A. Lidar$^{(2)}$, and Paolo Zanardi$%
^{(3)}$}
\affiliation{$^{(1)}$Institute of Theoretical Physics and Astrophysics, University of
Gd\'ansk, Poland\\
$^{(2)}$Departments of Chemistry, Electrical
Engineering-Systems, and Physics, University of Southern California, Los Angeles, CA 90089%
\\
$^{(3)}$Institute for Scientific Interchange (ISI), Villa Gualino, Viale
Settimio Severo 65, I-10133 Torino, Italy }

\begin{abstract}
We critically examine the internal consistency of a set of minimal
assumptions entering the theory of fault-tolerant quantum error correction
for Markovian noise. These assumptions are: fast gates, a constant
supply of fresh and cold ancillas, and a Markovian bath. We point out that these assumptions may not be mutually
consistent in light of rigorous formulations of the Markovian approximation.
Namely, Markovian dynamics requires either the singular coupling limit (high
temperature), or the weak coupling limit (weak system-bath interaction). The
former is incompatible with the assumption of a constant and fresh supply of
cold ancillas, while the latter is inconsistent with fast gates. We discuss
ways to resolve these inconsistencies.
As part of our discussion we derive, in the weak coupling limit,
a new master equation for a system subject to periodic driving. 
\end{abstract}

\maketitle

\section{Introduction}

The theory of fault-tolerant quantum error correction (FT-QEC) is one of the
pillars that the field of quantum information rests on. Starting with the
discovery of quantum error correcting codes \cite{Shor:95,Steane:96a}, and
the subsequent introduction of fault tolerance \cite{Shor:96}, this theory
has been the subject of many improvements and important progress \cite%
{Knill:96,Knill:98,Aharonov:96b,Aharonov:99,Zalka:96,Gottesman:97b,Gottesman:99,Gottesman:99a,Preskill:97a,Steane:99a,Steane:03,Knill:04,Reichardt:04,Knill:05}%
, leading to the well-known error correction threshold condition. Most
recently, work by Steane \cite{Steane:03} and Knill \cite{Knill:05} (see
also Reichardt \cite{Reichardt:04}) has pushed the threshold down to values
that are claimed to be very close to being within experimental reach. A
notable feature of much of the work on FT-QEC is that the error models are 
\emph{phenomenological}. By this we mean that the underlying models often do
not start from a Hamiltonian, microscopic description of the system-bath
interaction, but rather from a higher level, effective description, most
notably that of Markovian dynamics. E.g., Knill writes: \textquotedblleft We
assume that a gate's error consists of \emph{random, independent}
applications of products of Pauli operators with probabilities determined by
the gate\textquotedblright\ (our emphasis) \cite{Knill:05}. This approach is
natural given the considerable difficulty of obtaining error thresholds
starting from a purely Hamiltonian description. Nevertheless, Hamiltonian
approaches to decoherence management in a fault-tolerant setting have been
pursued, e.g., a mixed phenomenological-Hamiltonian treatment of FT-QEC \cite%
{MohseniLidar:05,Terhal:04,Aliferis:05,Aharonov:05}, and a fully Hamiltonian study of
fault tolerance in dynamical decoupling \cite{KhodjastehLidar:04}. Also
noteworthy are recent mixed phenomenological-Hamiltonian \emph{continuous
time} treatments of QEC \cite{Ahn:01,Sarovar:04,Sarovar:05}.

Here we are concerned with a critical re-evaluation of the physical
assumptions entering the theory of FT-QEC. We scrutinize, in particular, the
consistency of the assumption of Markovian dynamics within the larger
framework of FT-QEC. We point out that there may be an inherent
inconsistency in the theory of Markovian FT-QEC, when viewed from the
perspective of the validity of the Markovian approximation. We begin by
briefly reviewing, in Section~\ref{FTQEC-review}, a set of minimal and
standard, universally agreed upon assumptions made in Markovian FT-QEC theory. We then
review, in Section~\ref{MME-review}, the derivation of Markovian master
equations, emphasizing the physical assumptions entering the Markovian
approximation, in particular the requirement of consistency with
thermodynamics. Having delineated the set of assumptions entering FT-QEC and
the quantum Markov approximation, we discuss in Section~\ref{consistency}
the internal consistency of Markovian FT-QEC theory. We point out where
according to our analysis there is an inconsistency, and discuss possible
objections. In Section~\ref{alternatives} we then discuss how one may
overcome the inconsistency using a variety of alternative approaches,
including adiabatic quantum computing (QC), holonomic QC, topological QC,
and recent work on FT-QEC in a non-Markovian setting \cite%
{Terhal:04,Aliferis:05,Aharonov:05}. We conclude in Section~\ref{conc}.

\section{Review of Standard Assumptions of FT-QEC}

\label{FTQEC-review}

The following are a set of minimal assumptions made in the theory of FT-QEC 
\cite%
{Shor:96,Knill:96,Knill:98,Aharonov:96b,Aharonov:99,Zalka:96,Gottesman:97b,Gottesman:99,Gottesman:99a,Preskill:97a,Steane:99a,Steane:03,Knill:04,Reichardt:04,Knill:05}%
:

\begin{enumerate}
\item \textbf{A1} --- \emph{Gates can be executed in a time} $\tau _{g}$ 
\emph{such that} $\tau _{g}\omega =O(\pi )$, \emph{where} $\omega $ \emph{is
a typical Bohr or Rabi frequency}.\footnote{%
One might object that slower (even adiabatic) gates could be used instead.
We analyze this possibility in detail in Section~\ref{adiabatic-gates}, and
show that it does not lead to an improvement.}

\item \textbf{A2} --- \emph{A constant supply of fresh and nearly pure
ancillas}: at every time step we are given a supply of many qubits in the
state $|0\rangle $, each of which can be faulty with some error parameter $%
\eta \ll 1$.

\item \textbf{A3} --- \emph{Error correlations decay exponentially in time
and space}.
\end{enumerate}

Some remarks:

\noindent (i) \textbf{A1} is not typically stated explicitly in the
FT-QEC\ literature, but can be understood as resulting from the definition
of a quantum gate, which is a unitary transformation $U=\exp (iA)$; when $%
A=\tau _{g}H$, where $H$ is a Hamiltonian generating the gate, \textbf{A1}
follows from the absence of a free parameter:\ when $\tau _{g}$ is scaled up 
$H$ (and hence its eigenvalues) must be scaled down, and \textit{vice versa}.

\noindent (ii) The distinction between Bohr and Rabi frequencies in \textbf{%
A1} is related to the application of constant \textit{vs} periodic control
fields, respectively. In the case of a constant control field \textbf{A1}
can be understood as the condition that saturates the \textquotedblleft
Margolus-Levitin theorem\textquotedblright\ \cite{Margolus:98}, which states
that the time required to transform an initial state $|\psi \rangle $ to an
orthogonal state $|\psi ^{\bot }\rangle $ using a constant Hamiltonian $H$
is lower-bounded by $\tau _{\min }=\pi \hbar /(2E)$, where $E=\langle \psi
|H|\psi \rangle $; when $|\psi \rangle $ is an eigenstate of $H$ we have $%
\tau _{g}\geq \pi /(2\omega )$, where $\omega =E/\hbar $ is the Bohr
frequency. See also \cite{Andrecut:04} for the adiabatic version of the
Margolus-Levitin theorem, and \cite{Gea:02} for a lower bound on the amount
of energy needed to carry out an elementary logical operation on a quantum
computer, with a given accuracy and in a given time. In the case of periodic
control fields one can understand \textbf{A1} as the result of the standard
solution to the driven two-level atom problem, where the probability of a
transition between ground and excited state is given by $(\Omega _{\mathrm{R}%
}/\Omega _{\mathrm{R}}^{\prime })\sin ^{2}(\Omega _{\mathrm{R}}^{\prime
}t/2) $, where $\Omega _{\mathrm{R}}$ is the Rabi frequency and $\Omega _{%
\mathrm{R}}^{\prime }=(\Omega _{\mathrm{R}}+\delta ^{2})^{1/2}$, where $%
\delta $ is the detuning. This expression for the transition probability
yields $\tau _{g}\Omega _{\mathrm{R}}^{\prime }=O(\pi )$.

\noindent (iii) \textbf{A2} is shown to be necessary in \cite%
{Aharonov:96b}. \textbf{A3} is stated clearly in \cite{Aharonov:99} (see the
discussion in Sections 2.10 and 10 there). These, and additional assumptions
[such as constant fault rate (independent of number of qubits) and
parallelism (to correct errors in all blocks simultaneously)] are explicitly
listed, e.g., also in \cite{Dennis:02}, Section II.

\noindent (iv) \textbf{A3} is usually related to the
Markovian assumption, however both notions, the space-time correlations of
errors and the Markovian property, need some comments and explanations.
Using the convolutionless formalism in the theory of open systems (see, e.g.,
\cite{Breuer:02}) it is always possible to resolve the total superoperator $%
\Lambda (t)$ as 
\begin{equation}
\Lambda (t)=\prod_{i=1}^{n}\Lambda _{i}U_{i}  \label{eq:Lambda}
\end{equation}%
where $U_{i}$ are ideal unitary superoperators (corresponding to quantum
logic gates), and $\Lambda _{i}$ are linear maps, not necessarily completely
positive (CP) or even positive. If $\Lambda _{i}$ are CP then we can always
realize them by coupling to an evironment which is \textquotedblleft renewed
each time step\textquotedblright . This is the \textquotedblleft Markovian
condition\textquotedblright\ as formulated in \cite{Aharonov:99} (section
2.10). However, complete positivity is not a necessary condition for QEC,
which only requires a linear structure \cite{Knill:97b,ShabaniLidar:06}. To obtain the
Threshold Theorem one needs the following bound on the probability \cite%
{Aharonov:99}[Eq.~(2.6)]: 
\begin{equation}
\text{Pr}(\text{fault\ path\ with\ }k\text{\ errors})\leq c\eta ^{k}(1-\eta
)^{v-k},  \label{Pr}
\end{equation}%
where $\eta $ is the probability of a single error, $c$ is a certain
constant independent of $\eta $, and $v$ is the number of error locations in
the circuit. This bound implies that the $k$-qubit errors should scale as $%
\sim \eta ^{k}$, i.e., that in the decomposition of $\Lambda _{j}$ into $k$%
-qubit superoperators $L_{j}(k)$ 
\begin{equation}
\Vert L_{j}(k)\Vert \leq c\eta ^{k}\ .  \label{k-qub}
\end{equation}%
As discussed in \cite{Alicki:02} (within the Born approximation), the
condition (\ref{k-qub}) can strictly be satisfied only for temporally \emph{%
exponentially}{\ decaying reservoir correlation functions,} while for
realistic reservoir models the temporal decay is generically powerlike. The
decay of reservoir correlation functions (i.e., localization in time)
translates into localization of errors in space due to the finite speed of
error propagation. On the other hand it is widely believed that the
Markovian model can be understood as arising, to an excellent approximation,
from coupling to a reservoir which is not only renewed at each time step,
but whose influence is independent of the actual Hamiltonian dynamics of the
open system, and is localized in space (independent errors model) \cite%
{Giulini:book}. A large part of the present paper is devoted to a critical
discussion of this claim.

\noindent (v) We note that the recent papers on FT-QEC theory \cite%
{Terhal:04,Aliferis:05,Aharonov:05} relax the (Markovian) assumption \textbf{A3}, but do
make \textbf{A1} (implicitly) and \textbf{A2}. We comment on these
papers in Section~\ref{nonM-FTQEC}.

\section{Review of Markovian Master Equations}

\label{MME-review}

The field of derivations of the quantum Markovian master equation (MME) is
strewn with pitfalls: it is in fact non-trivial to derive the MME in a fully
consistent manner. There are essentially two types of fully rigorous
approaches, known as the \emph{singular coupling limit} (SCL) and the \emph{%
weak coupling limit} (WCL), both of which we consider below. See, e.g., the books 
\cite{Alicki:87,Breuer:book} for more details, as well as the derivation in 
\cite{Alicki:89}.

Consider a system and a reservoir (bath), with self Hamiltonians $H_{S}^{0}$
and $H_{R}$, interacting via the Hamiltonian $H_{SR}=\lambda S\otimes R$,
where $S$ ($R$) is a Hermitian system (reservoir) operator and $\lambda $ is
the coupling strength. A more general model of the form $H_{SR}=\sum_{\alpha
}\lambda _{\alpha }S_{\alpha }\otimes R_{\alpha }$ can of course also be
considered and results in the same qualitative conclusions. Thus the total
Hamiltonian is 
\begin{equation}
H=(H_{S}^{0}+H_{C}(t))\otimes I_{R}+I_{S}\otimes H_{R}+H_{SR},  \label{ham}
\end{equation}%
where $H_{C}(t)$ describes control over the quantum device (system), and $I$
is the identity operator.

The SCL and WCL derivations start from the expansion of the propagator $%
\Lambda $ for the reduced, system-only dynamics, 
\begin{equation}
\rho _{S}(t)=\Lambda (t,0)\rho _{S}(0),
\end{equation}%
computed in the interaction picture with respect to the renormalized, \emph{%
physical}, time-dependent Hamiltonian $H_{S}(t)=H_{S}+H_{C}(t)$, where 
\begin{equation}
H_{S}=H_{S}^{0}+\lambda ^{2}H_{1}^{\mathrm{corr}}(t)+\cdots .  \label{eq:H_S}
\end{equation}%
The renormalizing terms containing powers of $\lambda $ are
\textquotedblleft Lamb-shift\textquotedblright\ corrections due to the
interaction with the bath (see, e.g., \cite{Lidar:CP01}). The lowest order
(Born) approximation with respect to the coupling constant $\lambda $ yields 
$H_{1}^{\mathrm{corr}}$, while the higher order terms ($\cdots $) require
going beyond the Born approximation. Introducing a cumulant expansion for
the propagator, 
\begin{equation}
\Lambda (t,0)=\exp \sum_{n=1}^{\infty }[\lambda ^{n}K^{(n)}(t)],
\end{equation}%
one finds that $K^{(1)}=0$. The Born approximation consists of terminating
the cumulant expansion at $n=2$, whence we denote $K^{(2)}\equiv K$: 
\begin{equation}
\Lambda (t,0)=\exp [\lambda ^{2}K(t)+O(\lambda ^{3})].
\end{equation}%
One obtains%
\begin{eqnarray}
K(t)\rho _{S}=\int_{0}^{t}ds\int_{0}^{t}duF(s-u)S(s)\rho
_{S}S(u)^{\dag } \nonumber \\
+(\mathrm{similar \ terms})
\label{eq:K(t)}
\end{eqnarray}
as the first term in a cumulant expansion \cite{Alicki:89}. Here $F(s)=%
\mathrm{Tr}(\rho _{R}R(s)R)$ is the autocorrelation function, where $\rho
_{R}$ is the reservoir state and $R(s)$ is $R$ in the $H_{R}$-interaction
picture, and $S(u)$ is $S$ in the interaction picture with respect to the
physical Hamiltonian $H_{S}(t)$. The \textquotedblleft similar
terms\textquotedblright\ in Eq.~(\ref{eq:K(t)}) are of the form $\rho
_{S}S(s)S(u)^{\dag }$ and $S(s)S(u)^{\dag }\rho _{S}$.

At first sight $K(t)\sim t^{2}$, and this is true for small times (Zeno
effect \cite{Facchi:PRL02}). The Markov approximation means that we can
replace $K(t)$ by an expression that is linear in $t$, i.e. 
\begin{equation}
K(t)\simeq \int_{0}^{t}\mathcal{L}(s)ds  \label{eq:L}
\end{equation}%
where $\mathcal{L}(t)$ is a time-dependent Lindblad generator. That the
Lindblad generator can be time-dependent even after transforming back to the
Schr\"{o}dinger picture is important for our considerations below.

\subsection{Singular Coupling Limit}

\label{SCL}

The SCL approach we present in this subsection underlies the standard
derivation of the MME that can be found in almost any text concerning the
Markov approximation, though not always under the heading \textquotedblleft
SCL\textquotedblright\ (e.g., \cite{Carmichael:book}, p.8, Eq.~(1.36)). The
rigorous derivation of the SCL is briefly discussed (with references) in 
\cite{Alicki:87}, pp.36-38. It is based on a rescaling of the bath and
system-bath Hamiltonians, which physically makes sense in the
high-temperature limit only. We will shortly see the emergence of this limit.

In essence, the \textquotedblleft naive SCL-Markov
approximation\textquotedblright\ is obtained by the ansatz $F(s)=a\delta (s)$
for the autocorrelation function, whence 
\begin{equation}
L(s)\rho _{S}=aS(s)\rho _{S}S(s)^{\dag }+(\mathrm{similar\ terms}).
\end{equation}%
\textbf{\ }As a consequence, return to the Schr\"{o}dinger picture gives a
MME with the dissipative part independent of the Hamiltonian: 
\begin{eqnarray}
\frac{d\rho _{S}}{dt} &=&-i[H_{S}(t),\rho _{S}]+\mathcal{L}\rho _{S},  \notag
\\
\mathcal{L}\rho _{S} &\equiv &-\frac{1}{2}\lambda ^{2}a[S,[S,\rho _{S}]]
\label{eq:SCL}
\end{eqnarray}

More precisely, we must consider the multi-time bath correlation functions $%
F(t_{1},...,t_{n}):=\mathrm{Tr}[\rho _{R}R(t_{1})...R(t_{n})]:=\langle
R(t_{1})...R(t_{n})\rangle $. Here $R(t):=\exp (iH_{R}t)R\exp (-iH_{R}t)$
are the bath operators in the interaction picture, $\rho _{R}=\exp (-\beta
H_{R})/Z$ (where $\beta =1/kT$, $Z=\mathrm{Tr}[\exp (-H_{R}/kT)]$) is
the bath thermal equilibrium state at temperature $T$, which is a stationary
state of the reservoir, i.e., $[H_{R},\rho _{R}]=0$. The influence of the
environment on the system is entirely encoded into the $\{F(t_{1},...,t_{n})%
\}_{n=2}^{\infty }$.\footnote{$F(t_{1})$ is constant by stationarity. We
reserve the notation $F(t)$ for $F(t_{1},t_{2})\equiv F(t_{2}-t_{1})$ below.}
Heuristically, the Markov approximation can be justified under the following
conditions:

\begin{enumerate}
\item The lowest order correlation function, 
\begin{equation}
F(t)=\langle R(s+t)R(s)\rangle =\int_{-\infty }^{\infty }G(\omega
)e^{-i\omega t}d\omega ,  \label{2corr}
\end{equation}%
can be approximated by a Dirac delta function:\footnote{%
Note that stationarity implies that $F(t)$ does not depend on $s$.} 
\begin{equation}
F(t)\simeq \left( \int_{-\infty }^{\infty }F(s)ds\right) \delta
(t)=G(0)\delta (t)  \label{F-delta}
\end{equation}%
(white-noise approximation). Eq.~(\ref{2corr}) defines the \emph{spectral
density} $G(\omega )$, which is a key object in the theory.

\item Higher order correlation functions exhibit a Gaussian-type behavior,
i.e., can be estimated by sums of products of the lowest order ones, and
then, by condition (\ref{F-delta}), decay sufficiently rapidly.
\end{enumerate}

Let us now comment on the physical relevance of the white-noise
approximation.

First, the condition (\ref{F-delta}) cannot be satisfied in general. For
example, in the important case of linear coupling to a bosonic field (e.g.,
electromagnetic field, phonons in solid state), we have $G(0)=0$, which
means (by inverse Fourier transform) that $\int_{-\infty }^{+\infty
}F(t)dt=0 $, and therefore \emph{$F(t)$ cannot be well approximated by} $%
\delta (t)$.

Second, even for models with $G(0)>0$ there exists a universal relation, the
so-called Kubo-Martin-Schwinger (KMS) condition, $\langle R(t)R(0)\rangle
=\langle R(0)R(t+i\beta )\rangle $, which is valid for all quantum systems
at thermal equilibrium. This implies: 
\begin{equation}
G(-\omega )=e^{-\beta \omega }G(\omega )\ .  \label{KMS}
\end{equation}%
(See, e.g., \cite{Alicki:87}[pp.90-91], \cite{Thirring:book}[pp.176-177], or 
\cite{Breuer:book}[p.137].) The fundamental importance of the KMS condition
is captured by the fact that it is necessary in order for thermodynamics to
hold. The KMS condition implies a strong asymmetry of the spectral density $%
G(\omega )$ for low $T$, where $T$ is measured relative to the presence of $%
kT$ energy scales in the bath, i.e., relative to the range where $G(\omega )$
is non-vanishing. The KMS condition is relevant to our discussion since we
make the reasonably minimalistic assumption that the reservoir (not the QC)
is in thermal equilibrium.\footnote{%
One may challenge the notion that the bath must always be in thermal
equilibrium. E.g., consider an atom in a microwave cavity, with the cavity
electromagnetic field initially in thermal equilibrium. Now suppose the atom
is driven and is coupled to the cavity electromagnetic field, which
therefore is no longer in equilibrium. However, is the internal
electromagnetic field the relevant environment, or is it the external one?
Clearly, the electromagnetic field inside the cavity is not a reservoir but
itself a part of the system. This is because: a) its spectrum is discrete,
b) its coupling to the atom (close to resonance) is enhanced. The reason
these considerations matter is because b) implies the strong coupling
regime, hence failure of the initial state tensor product structure
assumption, hence difficulties with the separation of the system from the
reservoir (dressed atom picture); a) implies that $F(t)$ is (quasi)-periodic
with short Poincar\'{e} recurrences, hence a strongly non-Markovian regime,
and thus associated difficulties for Markovian FT-QEC. On the other hand the
external electromagnetic field has a continuous spectrum and the state
product structure is easily satisfied, hence qualifies as a reservoir. This
example merely serves to illustrate accepted notions regarding the division
into well defined system and bath; for most practical purposes a thermal
equlibrium is the simplest and most relevant model of an environment, and
FT-QEC\ theory must be applicable to this setting.}

Third, $G(\omega )$ need not be flat even at high $T$ (indeed, the KMS
condition only implies that $G(\omega )$ is symmetric at high $T$). For
example, this is the case for the electromagnetic field and for phonons, for
which at $T>0$ one has $G(\omega )\sim \omega ^{3}/(1-e^{-\hbar \omega /kT})$
for $|\omega |\leq \omega _{\mathrm{cut}}$, and $G(\omega )=0$ for $|\omega
|>\omega _{\mathrm{cut}}$. One can see that for high $T$ ($kT\gg \hbar
\omega _{\mathrm{cut}}$), $G(\omega )\sim kT\omega ^{2}$ is symmetric. Here $%
\omega _{\mathrm{cut}}$ is the Debye frequency in the case of phonons, while
for the electromagnetic field $\omega _{\mathrm{cut}}$ should tend to
infinity in the renormalization procedure. A flat $G(\omega )$ means a
structureless bath, while physical systems always have a nontrivial
structure depending on relevant energy scales.\footnote{%
It is interesting to note that even if we try to enforce a flat $G(\omega )$
by, e.g., choosing an appropriate form factor for the spin-boson system, the
obtained model -- the so-called \textquotedblleft Ohmic
case\textquotedblright\ -- is mathematically and physically ill defined (see 
\cite{Alicki:02a}).}

Now let us return to the implications of the SCL assumptions for the problem
of FT-QEC. In order to derive the SCL from first principles, one rescales $%
H_{R}\rightarrow H_{R}/\epsilon ^{2}$, rescales $H_{SR}\rightarrow
H_{SR}/\epsilon $, but keeps $H_{S}$ and $\rho _{R}$ fixed.\footnote{%
Note that because different Hamiltonians are rescaled differently, this
rescaling procedure is \emph{not} equivalent to a direct rescaling of the
time variable (which is what is done in the WCL, below).} The idea of this
rescaling is that it accelerates the reservoir's evolution (via $%
H_{R}\rightarrow H_{R}/\epsilon ^{2}$) and hence produces faster decay of
the reservoir correlations, $F(t)$. To see this, note that the rescaling $%
H_{SR}\rightarrow H_{SR}/\epsilon $ increases the amplitude $F(0)$ to $%
F(0)/\epsilon ^{2}$ (proportional to $H_{SR}^{2}$), while keeping the
strength of the noise $\int_{-\infty }^{+\infty }F(t)dt=G(0)$ fixed (as can
be seen via a change of variables $t\rightarrow t/\epsilon ^{2}$ in the
integral). This implies a faster decay of $F(t)$. The rescaling procedure is
specifically designed to yield the delta correlation [Eq.~(\ref{F-delta})]
in the limit as $\epsilon \rightarrow 0$. Note that if $\rho _{R}$ is at
thermal equilibrium at temperature $T$ with respect to $H_{R}$, then, since $%
\rho _{R}=\exp (-\beta H_{R})/Z$ is fixed, it must be at thermal equilibrium
with respect to $H_{R}/\epsilon ^{2}$ at the temperature $T/\epsilon
^{2}\rightarrow \infty $, whence our mention of the high temperature limit,
above. Further note that $H_{S}$ is not rescaled since the SCL is
(artificially) designed to produce \textquotedblleft white
noise\textquotedblright\ on the natural time scale of system's evolution,
which is given by $H_{S}$.

Another, equivalent way to understand the emergence of the high-$T$ limit is
the following: For the Markovian condition $F(t)\simeq a\delta (t)$ to hold
the spectral density must be flat: $G(\omega )=\mathrm{const}$. However,
this is possible only in the limit $T\rightarrow \infty $ of the KMS
condition. More precisely, \emph{the Markovian condition can hold only if }$%
kT\gg \omega $ \emph{over the entire spectrum of the system's Bohr
frequencies}. Strictly speaking, $G(\omega )$ is never constant. The
variation of $G(\omega )$ happens over the \textquotedblleft thermal
memory\textquotedblright\ time $\tau _{\mathrm{th}}:=1/kT$. In the infinite $%
T$ limit we then recover the case of zero memory-time, i.e., Markovian
dynamics. Physically, it is enough to assume that $G(\omega )$ is
essentially constant over the interval $[-\omega _{0},\omega _{0}]$ where $%
kT>\omega _{0}\gg $ system's Bohr frequencies. I.e., system energy scales
must be compared to $1/\tau _{\mathrm{th}}$ and this leads to the important
realization that \emph{the Markovian approximation can be consistent with
the KMS condition only in the high temperature regime }$kT\gg E$, \emph{where%
} $E$ \emph{is the system energy scale}. As we argue below, this fact
presents a serious difficulty in the context of FT-QEC, the issue being
essentially that the requirement of a constant supply of nearly pure and
cold ancillas contradicts the high-$T$ limit needed for the Markov
approximation to hold.

\subsection{Weak Coupling Limit}

\label{WCL}

In the SCL approach above there was no restriction on the time-dependence of
the system Hamiltonian. However, the price paid is the high-$T$ limit.
Moreover, while mathematically the SCL is rigorous in the scaling limit, it
is inconsistent with thermodynamics except in the $T\rightarrow \infty $
limit. On the other hand, the derivation by Davies, in his seminal 1974
paper \cite{Davies:74}, is perhaps the only derivation of the MME that is
entirely consistent from \emph{both the mathematical and physical} points of
view. The Davies approach is based not on the high-$T$ limit, but rather on
the physically plausible idea of weak coupling. This is natural and
consistent with thermodynamics at all temperatures.

More specifically, Davies' derivation does not invoke a flatness condition
on $G(\omega )$ but is, of course, still subject to the KMS condition. In
the Davies approach the Markov approximation is a consequence of weak
coupling (and hence slow dynamics of the system in the interaction picture),
and time coarse-graining, which leads to cancellation of the non-Markovian
oscillating terms. The price we pay is the invalidity of this approach for
time-dependent Hamiltonians, except in the adiabatic case. We explain this
important comment below. Hence, while the Davies approach does not require
the high-$T$ limit, it imposes severe restrictions on the speed of quantum
gates.

In his rigorous derivation Davies replaced the heuristic condition (\ref%
{F-delta}) by the weaker 
\begin{equation}
\int |F(t)|dt<\infty .  \label{F-cond}
\end{equation}%
This condition avoids the difficulties originating from the singularity of
the SCL condition (\ref{F-delta}), and preempts the corresponding problems
with the high-$T$ limit.\footnote{%
In some sense the weak coupling limit is similar to the Central Limit
Theorem (CLT) in probability, and condition~(\ref{F-cond}) is analogous to a
rough upper bound on the second moment in the CLT. If it is not satisfied
then the noise may be not Gaussian in the weak coupling limit. The value of $%
\int |F(t)|dt$ itself does not provide any meaningful physical parameter and
can depend on some regularization/cut-off parameters.}$^{,}$\footnote{%
One can go further and ask how generic the Markovian case is, in the sense
that Eq.~(\ref{F-cond}) is satisfied. In fact, typically $F(t)$ decays as $%
1/t^{\alpha }$ (e.g., for the vacuum bath $\alpha =4$ \cite{Alicki:02}%
\textbf{)}, which means that in some cases ($\alpha \leq 1$) Eq.~(\ref%
{F-cond}) can be violated. For a systematic treatment of these non-Markovian
effects see, e.g., \cite{Breuer:02}.} We now consider the cases of a
constant, periodic, and arbitrarily time-dependent control Hamiltonian. The
constant case is the one originally treated by Davies \cite{Davies:74}, and
extended in \cite{Davies:78} to time-dependent Hamiltonians assuming a slow
(\textquotedblleft adiabatic\textquotedblright ) change on the dissipation
time scale $\lambda ^{2}t$. The non-constant cases we study here have, as
far as we know, not been published before in the general scientific
literature.

\subsubsection{WCL for Constant $H_{C}$: Summary of the Original Davies
Derivation}

We present a simplified version of the discussion of the Markov
approximation in \cite{Alicki:89}. Denote by ${E_{k}}$ the Bohr energies
(eigenvalues of $H_{S}$), let $\omega \in \{\omega _{kl}=E_{k}-E_{l}\}_{k,l}$%
, and let $S_{\omega }$ be the discrete Fourier components of the
interaction picture $S$, i.e., 
\begin{equation}
S(t)=\exp (iH_{S}t)S\exp (-iH_{S}t)=\sum_{\omega }S_{\omega }\exp (i\omega
t),  \label{eq:S}
\end{equation}%
where $H_{S}$ is the renormalized (physical) system Hamiltonian: the sum of
the \textquotedblleft bare\textquotedblright\ $H_{S}^{0}$ and a Lamb shift
term (bath induced), as in Eq.~(\ref{eq:H_S}). Equivalently, 
\begin{equation}
\lbrack H_{S},S_{\omega }]=\omega S_{\omega }.  \label{Dav1}
\end{equation}%
We remark that in the original Davies paper the Bohr energies and Eq.~(\ref%
{Dav1}) are computed with respect to the bare Hamiltonian $H_{S}^{0}$. Here
we use the physical Hamiltonian $H_{S}$ in order to take into account the
fact that the Lamb shift term, although formally proportional to $\lambda
^{2}$, can be large or even infinite after cut-off removal.

Then, it follows from Eq.~(\ref{eq:K(t)}) that 
\begin{eqnarray}
K(t)\rho _{S}&=&\sum_{\omega ,\omega ^{\prime }}S_{\omega }\rho _{S}S_{\omega
^{\prime }}^{\dag }\int_{0}^{t}e^{i(\omega -\omega ^{\prime
  })u}du\int_{-u}^{t-u}F(\tau )e^{i\omega \tau }d\tau \notag \\
&+&(\mathrm{similar}\text{ 
}\mathrm{terms}).  \label{eq:K2}
\end{eqnarray}
The weak coupling limit is next formally introduced by rescaling the time $t$
to $t/\lambda ^{2}$ (van Hove limit). This enables two crucial
approximations, which are valid in the resulting large-$t$ limit:

\begin{enumerate}
\item We replace\footnote{%
In a more rigorous treatment the Cauchy principal value must be used, but
the result is essentially the same \cite{Alicki:89}.}%
\begin{equation}
\int_{0}^{t}e^{i(\omega -\omega ^{\prime })u}du\approx t\delta _{\omega
\omega ^{\prime }}.  \label{eq:rep1}
\end{equation}%
This makes sense for 
\begin{equation}
t\gg \max \{1/(\omega -\omega ^{\prime })\}.  \label{eq:tbig}
\end{equation}%
\emph{This violates \textbf{A1}, expressed in terms of the Bohr frequencies}%
. We see here already the emergence of an adiabatic criterion for the
validity of the Markov approximation.

\item We replace $\int_{-u}^{t-u}F(\tau )e^{i\omega \tau }d\tau $ by the
Fourier transform:%
\begin{equation}
\int_{-u}^{t-u}F(\tau )e^{i\omega \tau }d\tau \approx {G}(\omega
)=\int_{-\infty }^{\infty }F(\tau )e^{i\omega \tau }d\tau .  \label{eq:rep2}
\end{equation}%
\end{enumerate}

The physical validity of the last approximation is usually ignored, though one
can make the following argument: On the LHS of Eq.~(\ref{eq:rep2}), for a
given Bohr frequency $\omega $ the Fourier-like integral must sample the
function $F(\tau )$ with sufficiently high accuracy so that the Fourier
transform approximation will be valid. To this end one needs a time $t$ such
that: (i) $t\gg 1/\omega $. This is a weaker condition than the previous one
[$t\gg \max \{1/(\omega -\omega ^{\prime })\}$] which involves differences
of Bohr frequencies. (ii) The time $t$ must be also much longer than the
time scale of the wildest variations of $F(\tau )$, which is typically [as
may be checked for simple models of spectral densities $G(\omega )$] given
by $1/\omega _{\mathrm{cut}}$, where $\omega _{\mathrm{cut}}$ is a
high-frequency cutoff. When $\omega <\omega _{\mathrm{cut}}$ (i) implies
(ii). Therefore typically Eq.~(\ref{eq:rep1}) is a stronger assumption than
Eq.~(\ref{eq:rep2}).

Applying the approximations (\ref{eq:rep1}) and (\ref{eq:rep2}), we obtain $K(t)\rho _{S}=t\sum_{\omega
}S_{\omega }\rho _{S}S_{\omega }^{\dag }{G}(\omega )+(\mathrm{similar}$ $%
\mathrm{terms})$, and hence it follows from Eq.~(\ref{eq:L}) that $\mathcal{L%
}(s)=\mathcal{L}$ is the Davies generator in the familiar Lindblad form: 
\begin{eqnarray}
\frac{d\rho _{S}}{dt} &=&-i[H_{S},\rho _{S}]+\mathcal{L}\rho _{S},  \notag \\
\mathcal{L}\rho _{S} &\equiv &\frac{1}{2}\lambda ^{2}\sum_{\omega }G(\omega
)([S_{\omega },\rho S_{\omega }^{\dagger }]+[S_{\omega }\rho ,S_{\omega
}^{\dagger }])  \label{Dav}
\end{eqnarray}%
Several remarks are in order:

\noindent (i) The absence of off-diagonal terms in Eq.~(\ref{Dav}), compared
to Eq.~(\ref{eq:K2}), is the hallmark of the Markovian limit. Namely, the
Davies derivation relies on the cancellation of the non-Markovian
off-diagonal terms $\sum_{\omega \neq \omega ^{\prime }}S_{\omega }\rho
_{S}S_{\omega ^{\prime }}^{\dag }\int_{0}^{t}e^{i(\omega -\omega ^{\prime
})u}du$. This time coarse-graining is possible due to integration over the
fast-oscillating $\int_{0}^{t}e^{i(\omega -\omega ^{\prime })u}$ terms over
a long timescale, i.e., over $t\gg \max \{1/(\omega -\omega ^{\prime })\}$
(see also \cite{Lidar:CP01}). As remarked above, this violates \textbf{A1},
expressed in terms of the Bohr frequencies.

\noindent (ii) It follows from Bochner's theorem applied to the Fourier
transform definition of ${G}(\omega )$ that $G(\omega )\geq 0$ \cite%
{Alicki:87}[p.90], \cite{Breuer:book}[p.136]; this result is essential for
the complete positivity of the Markovian master equation in the WCL.

\noindent (iii) Davies' derivation showed implicitly that the notion of
\textquotedblleft bath's correlation time\textquotedblright\ is not
well-defined -- Markovian behavior involves a rather complicated cooperation
between system and bath dynamics. More specifically, the relations~(\ref{Dav}%
) and~(\ref{Dav1}) together imply that the noise and $H_{S}$ are strongly
correlated. In other words, contrary to what is often done in
phenomenological treatments, \emph{one cannot combine arbitrary }$H_{S}$%
\emph{'s with given }$S_{\omega }$\emph{'s.} This point is particularly
relevant in the context of FT-QEC, where it is common to assume Markovian
dynamics and apply arbitrary control Hamiltonians.

Davies did not consider time-dependent system Hamiltonians in \cite%
{Davies:74}, but it is possible to generalize his derivation to allow for
slowly varying system Hamiltonians \cite{Davies:78,Alicki:79,Alicki:89}.
That is, whenever the time scale of the variation of $H_{C}(t)$ is much
longer than the inverse of the typical Bohr frequency (of $H_{S}$), it is
possible to add $H_{C}(t)$ to the system Hamiltonian in Eq.~(\ref{Dav}),
necessitating at the same time this change also in Eq.~(\ref{Dav1}). This is
a type of adiabatic limit (indeed, the $S_{\omega }$ in Eq.~(\ref{Dav1}) can
be interpreted, with $H_{S}$ replaced by $H_{S}+H_{C}(t)$, as being
adiabatic eigenvectors of the superoperator $[H_{S}+H_{C}(t),\cdot
]$). We note that an alternative approach to adiabaticity in open
quantum systems was recently developed in
Ref.~\cite{SarandyLidar:04}. This approach, while being very general, is more
phenomenological in that it postulates a
convolutionless master equation, and then derives corresponding
adiabaticity conditions. Closer in spirit to the Davies derivation is
another recent approach to adiabaticity in open systems, which assumes
slow system variation together with weak system-bath coupling \cite{Thunstrom:05}.

\subsubsection{WCL for Periodic Driving:\ Floquet Analysis}

\label{Floquet}

Before considering the case of periodic $H_{C}$ let us consider briefly once
more the case of a constant Hamiltonian in the so-called covariant
dissipation setting. Covariance is the commutation condition $\mathcal{H}%
\mathcal{L}=\mathcal{L}\mathcal{H}$ where $\mathcal{H}=[H_{S},\cdot ]$ is
the super-operator constant Hamiltonian, and $\mathcal{L}$ is the Davies
generator. Covariance is an abstract property which is automatically
fulfilled for the Davies generator.\footnote{%
This can be verified by directly computing $\mathcal{HL}$ and making use of
Eq.~(\ref{Dav1}) and the relation $[A,BC]=[A,B]C+B[A,C]$ (for operators $A,B$
and $C$). A more elegant way to see this is to consider $\mathcal{L}(t)=\exp
(-it\mathcal{H})\mathcal{L}\exp (it\mathcal{H})$ and note that Eq.~(\ref%
{Dav1}) implies that $S(t)$ and $S^{\dag }(t)$ rotate in opposite
directions. Hence $\mathcal{L}(t)=\mathcal{L}$, whence $d\mathcal{L}(t)/dt=0$
gives the result.} It is convenient since it implies factorization of the
full propagator into Hamiltonian and dissipative parts. Markovian dynamics
obtained in the WCL as discussed above takes the form

\begin{equation}
{\frac{d\rho }{dt}}=(-i\mathcal{H}+\mathcal{L})\rho \ ,\ t\geq 0,  \label{1}
\end{equation}%
where the most general form of the Lindblad (or Davies) $\mathcal{L}$
satisfying Eq.~(\ref{1}) is 
\begin{equation}
\mathcal{L}\rho ={\frac{1}{2}}\sum_{\{\omega \},j}\left( [V_{j}(\omega
),\rho V_{j}(\omega )^{\dagger }]+[V_{j}(\omega )\rho ,V_{j}(\omega
)^{\dagger }]\right) .  \label{2}
\end{equation}%
Here $\{\omega \}\equiv \mathrm{Spectrum}(\mathcal{H})$, i.e., the Bohr
frequencies (differences of eigenvalues of $H$), and 
\begin{equation}
\mathcal{H}V_{j}(\omega )=\omega V_{j}(\omega )  \label{3}
\end{equation}%
[i.e., Eq.~(\ref{Dav1})]. The solution, i.e., the dynamical semigroup is%
\begin{equation}
\rho (t)=\Lambda (t)\rho (0),\ \Lambda (t)=e^{-it\mathcal{H}}e^{t\mathcal{L}%
}.  \label{4}
\end{equation}%
Now consider a periodic control Hamiltonian with period $\Theta $ 
\begin{equation}
H_{C}(t)=H_{C}(t+\Theta ),\ \Omega =2\pi /\Theta .  \label{5}
\end{equation}%
(Note that $\Omega $ is \emph{not} the Rabi frequency, which throughout this
paper we denote by $\Omega _{\mathrm{R}}$.) The situation is then very
similar to the standard (time-independent $H_{C}$) WCL, but the set of
\textquotedblleft effective Bohr frequencies\textquotedblright\ (Floquet
spectrum)\ ${\omega }$ is now larger and is of the form $\{\omega +q\Omega
\} $, $q=0,\pm 1,...$. Here $\omega $ are Bohr frequencies for the Floquet
unitary [defined in Eq.~(\ref{eq:floquet}) below], i.e., differences of
eigenvalues $\epsilon _{\alpha }$ of the Floquet unitary, rather than $%
\{\omega \}=\mathrm{Spectrum}(\mathcal{H})$ as above. As this set of
\textquotedblleft effective Bohr frequencies\textquotedblright\ is discrete
the WCL still works, but the final Davies generator is more complicated, as
we now show.

Define the time-ordered unitary propagator 
\begin{equation}
U(t,s)\equiv \mathcal{T}\exp \left( -i\int_{s}^{t}H_{S}(u)du\right) ,\ t\geq
s  \label{6}
\end{equation}%
which satisfies the properties $U(s,t)\equiv U(t,s)^{-1}=U(t,s)^{\dagger }$, 
$U(t,s)U(s,u)=U(t,u)$, $U(t+\Theta ,s+\Theta )=U(t,s)$, and ${\frac{d}{dt}}%
U(t,s)=-iH_{S}(t)U(t,s)$, ${\frac{d}{dt}}U(t,s)^{\dagger }=iU(t,s)^{\dagger
}H_{S}(t)$. The \emph{Floquet unitary operator} is%
\begin{equation}
F(s)\equiv U(s+\Theta ,s)\bigskip ,  \label{eq:floquet}
\end{equation}%
with corresponding super-operator action%
\begin{equation}
\mathcal{F}(s)\rho \equiv F(s)\rho F(s)^{\dagger },
\end{equation}%
and Floquet eigenvectors $|\phi _{\alpha }\rangle $ and eigenvalues
(quasi-energies) $\epsilon _{\alpha }$ satisfying\footnote{%
Note that the Floquet Hamiltonian $H_{S}(t)-id/dt$ operates on a different
Hilbert space than $F(0)$ (the space of periodic functions with values in
the system's Hilbert space). But its eigenvalues coincide with $\epsilon
_{\alpha }$ from Eq.~(\ref{19}).} 
\begin{equation}
F(0)|\phi _{\alpha }\rangle =e^{-i\epsilon _{\alpha }\Theta }|\phi _{\alpha
}\rangle .  \label{19}
\end{equation}%
It follows from standard Floquet theory that%
\begin{equation}
U(t,0)|\phi _{\alpha }\rangle =e^{-it\epsilon _{\alpha }}\sum_{q\in \mathbf{Z%
}}e^{-itq\Omega }|\phi _{\alpha }(q)\rangle ,  \label{20}
\end{equation}%
i.e., the set $\{|\phi _{\alpha }(q)\rangle \}$ is a complete basis.
Therefore we have at most as many $q$'s as the dimension of the Hilbert
space. That the number of $q$'s is finite is important for our
considerations below.

We call a Lindblad generator $\mathcal{L}$ a \textquotedblleft covariant
dissipative perturbation of $H_{S}(t)$\textquotedblright\ if 
\begin{equation}
\mathcal{F}(0)\mathcal{L}=\mathcal{L}\mathcal{F}(0)  \label{9}
\end{equation}

We will assume this property, similarly to the case of a constant
Hamiltonian described above. In fact, covariance holds for a periodic $%
H_{S}(t)$ and also for the corresponding WCL Davies generator. One can then
derive the \emph{covariant master equation} (we sketch the
derivation below):%
\begin{equation}
{\frac{d\rho }{dt}}=\left( -i\mathcal{H}(t)+\mathcal{L}(t)\right) \rho \ ,\
t\geq 0,  \label{10}
\end{equation}%
[compare to Eq.~(\ref{1})] where 
\begin{eqnarray}
\mathcal{L}(t) &=&\mathcal{U}(t,0)\mathcal{L}\mathcal{U}(t,0)^{\dagger },
\label{11} \\
{\frac{d}{dt}}\mathcal{U}(t,s) &=&-i\mathcal{H}(t)\mathcal{U}(t,s),
\end{eqnarray}%
and where the general form of $\mathcal{L}$ appearing in Eq.~(\ref{11}) is
given by Eq.~(\ref{2}), with $V_{j}(\omega )$ now being eigenvectors of $%
\mathcal{F}(0)$, 
\begin{equation}
\mathcal{F}(0)V_{j}(\omega )=e^{-i\omega \Theta }V_{j}(\omega ),  \label{14}
\end{equation}%
rather than of $\mathcal{H}$, as in Eq.~(\ref{3}). Moreover, here $\{\omega
\}\equiv \{\epsilon _{\alpha }-\epsilon _{\beta }\}$, where $\epsilon
_{\alpha }$ are quasi-energies (effective Bohr frequencies) of the Floquet
operator, rather $\{\omega \}\equiv \mathrm{Spectrum}(\mathcal{H})$ as we
saw in the case of constant $H_{C}$.

The solution replacing Eq.~(\ref{4}) is 
\begin{eqnarray}
\rho (t) &=&\Lambda (t,s)\rho (s),\text{ \ }t\geq s  \notag \\
\Lambda (t,s) &=&\mathcal{T}\exp \left\{ \int_{s}^{t}\left( -i\mathcal{H}(u)+%
\mathcal{L}(u)\right) du\right\}  \label{12}
\end{eqnarray}%
By direct computation one can prove the following properties: 
\begin{eqnarray}
\mathcal{L}(t+\Theta ) &=&\mathcal{L}(t), \\
\mathcal{F}(s)\mathcal{L}(s)\mathcal{F}(s)^{\dagger } &=&\mathcal{L}(s), \\
\Lambda (t,s)\Lambda (s,u) &=&\Lambda (t,u)\text{ for }t\geq s\geq u, \\
\Lambda (t+\Theta ,s+\Theta ) &=&\Lambda (t,s), \\
\Lambda (t,s) &=&\mathcal{U}(t,s)e^{-(t-s)\mathcal{L}(s)}.  \label{13}
\end{eqnarray}

To derive the covariant master equation (\ref{10}) one considers the
standard picture of an open system $S+R$ with the total Hamiltonian 
\begin{equation}
H_{SR}(t)=H_{S}^{0}(t)+H_{R}+\sum_{k}S_{k}\otimes R_{k},  \label{15}
\end{equation}%
(we neglect the Lamb shift correction here; it can be included, changing $%
H_{S}^{0}(t)$ into the physical Hamiltonian $H_{S}(t)$, by a suitable
renormalization procedure), stationary reservoir state $\rho _{R}$, $%
[H_{R},\rho _{R}]=0$, $\mathrm{Tr}(\rho _{R}\cdot )\equiv \langle \cdot
\rangle _{R}$, $\langle R_{k}\rangle _{R}=0$. Then, exactly following a
Davies-like calculation using a Fourier decomposition of $S(t)$, now
governed by a periodic Hamiltonian, and making in particular again the
crucial assumption Eq.~(\ref{eq:tbig}), which now reads 
\begin{equation}
t\gg \max \{1/(\omega -\omega ^{\prime }+m\Omega )\},~m=0,\pm 1,\pm 2,...
\label{MA-per}
\end{equation}%
with $|m|$ upper-bounded by the dimension of the Hilbert space [see remark
after Eq.~(\ref{20})], one obtains Eq.~(\ref{10}) in the Davies WCL. The
explicit form of the generator is: 
\begin{eqnarray}
\mathcal{L}\rho &=& {\frac{1}{2}}\sum_{k,l}\sum_{q\in \mathbf{Z}}\sum_{\{\omega
\}}{\hat{R}}_{kl}(\omega +q\Omega )\{ [S_{l}(q,\omega )\rho
  ,S_{k}(q,\omega )^{\dagger }] \nonumber \\
&+& [S_{l}(q,\omega ),\rho S_{k}(q,\omega
)^{\dagger }] \} .  \label{16}
\end{eqnarray}
Here $\{\omega \}\equiv \{\epsilon _{\alpha }-\epsilon _{\beta }\}$, the
Floquet spectrum, and 
\begin{equation}
{\hat{R}}_{kl}(x)=\int_{-\infty }^{\infty }e^{-itx}\langle
R_{k}(t)R_{l}\rangle _{R}dt  \label{17}
\end{equation}%
\begin{equation}
S_{k}(q,\omega )=\sum_{p\in \mathbf{Z}}\sum_{\{\epsilon _{\alpha }-\epsilon
_{\alpha ^{\prime }}=\omega \}}\langle \phi _{\alpha }(p+q)|S_{k}|\phi
_{\alpha ^{\prime }}(p)\rangle |\phi _{\alpha }\rangle \langle \phi _{\alpha
^{\prime }}|.  \label{18}
\end{equation}%
$S_{k}(q,\omega )$ is the part of $S(t)$ which rotates with frequency $%
\omega +q\Omega $ and can be computed using Eq.~(\ref{20}). Note that by
diagonalizing the matrices ${\hat{R}}_{kl}$ one can transform the generator $%
\mathcal{L}$ of Eq.~(\ref{16}) into the form of Eq.~(\ref{2}), which allows
one to read off the operators $V_{j}(\omega )\ $appearing there.

Now to some important comments:

\noindent -- \emph{Timescale analysis}: Note that for the periodic case the
differences of \textquotedblleft Bohr frequencies\textquotedblright\ may be
of the order of $1/\Theta $. Hence we conclude from Eq.~(\ref{MA-per})\ that
one must average over many periods $\Theta $, i.e., require $t\gg \Theta $.
This can be interpreted as a condition that \textquotedblleft the
environment must learn that the Hamiltonian is periodic\textquotedblright .
This is exactly analogous to the adiabaticity condition in the adiabatic
case: $H(t)$ must be constant over many inverse Bohr frequencies to
\textquotedblleft be recognised\textquotedblright\ by the environment. The
periodic WCL is also a coarse-grained time description with the additional
time scale $\Theta $.
\emph{Note that arbitrarily fast
periodic driving (small $\Theta $) is incompatible even with the kind
of generalized, finitely localized MME derived here, since then differences
of Bohr frequencies matter in Eq.~(\ref{MA-per})} (recall that $\max
|m|$ is bounded by the -- typically small -- dimension of the system
Hilbert space).\\

\noindent -- \emph{Where is the Rabi frequency?} Note the dependence of the
operators $S_{k}(q,\omega )$ on the Floquet eigenvalue differences $\epsilon
_{\alpha }-\epsilon _{\alpha ^{\prime }}$. The usual Rabi frequency, $\Omega
_{\mathrm{R}}=2dE/\hbar $ ($d$ is the dipole moment, $E$ is the electric
field amplitude) arises in the dipole approximation, which we have not made
here. The usual Rabi frequency is replaced in our non-perturbative treatment
(in the sense of no multipole expansion) by the difference of Floquet
eigenvalues $\epsilon _{\alpha }-\epsilon _{\alpha ^{\prime }}$ in Eq.~(\ref%
{18}).\footnote{%
One can see that such a term also arises in the usual dipole approximation
by considering, e.g., Eq. (2.94) in \cite{Carmichael:book}. The interaction
picture raising and lowering operators $\sigma _{\pm }(t)$ (for a two-level
atom driven by a classical field) there oscillate with three
\textquotedblleft Bohr frequencies\textquotedblright\ $\omega _{A},\omega
_{A}\pm \Omega $\thinspace , where $\Omega =2dE/\hbar $ denotes the usual
Rabi frequency. Hence the Rabi frequency is a difference of two Bohr
frequencies.}

\noindent -- \emph{More on the Rabi frequency}:\ As we saw, the non-Markovian terms
vanish because of the time coarse-grained description. To attain this, we
must average over times $t$ $\gg \max_{\omega ,\omega ^{\prime }}\{1/(\omega
-\omega ^{\prime })\}$, but must also keep in mind that the longest relevant
scale for coarse-graining is given by the exponential decay time $\tau $ (a 
\emph{derived} quantity), i.e., we must have $t<\tau $. The Rabi frequency $%
\Omega _{\mathrm{R}}$ is a difference of two Bohr frequencies $\omega
,\omega ^{\prime }$:\ $\Omega _{\mathrm{R}}=\omega -\omega ^{\prime }$. This
implies that coarse-graining does not makes sense if $\Omega _{\mathrm{R}%
}\tau \ll 1$ [since then $t<\tau \ll 1/\Omega _{\mathrm{R}}=1/(\omega
-\omega ^{\prime })$, in contradiction to the fundamental requirement on $t$%
]. In physical terms this means that the width of the spectral line ($\gamma
=1/\tau $) is larger than the level splitting $\Omega _{\mathrm{R}}$ (see,
e.g., Fig. 2.5 (i),(ii) in \cite{Carmichael:book} for an illustration in the
case of the incoherent fluorescence spectrum) and therefore
\textquotedblleft the environment has no time to recognize the details of
the spectrum\textquotedblright . On the other hand, when $\Omega _{\mathrm{R}%
}\tau \gg 1$ (not inconsistent with the WCL), $\Omega _{\mathrm{R}}$ must
appear in the generator, as appears from our treatment of the case of
periodic driving in the WCL, above. Unfortunately there are examples in the
literature where an MME\ is written down subject to $\Omega _{\mathrm{R}%
}\tau \gg 1$ but $\Omega _{\mathrm{R}}$ does not appear in the generator
[e.g., Eq.~(2.96) in \cite{Carmichael:book}, where $\Omega _{\mathrm{R}}\sim
10^{10}\mathrm{Hz}$ and $\tau \sim 10^{-8}\mathrm{s}$].

\noindent -- \emph{Quantum optics considerations}:\ The Markov approximation is
commonly accepted as an excellent approximation in quantum optics; see,
e.g., the discussion of resonance fluorescence in \cite{Carmichael:book}%
[Ch.2]. This is also the basis for substantial confidence in the possibility
of FT-QEC in quantum optical systems, such as trapped ions \cite{Cirac:95}
and atoms trapped in microwave cavities \cite{Turchette:95}. Such arguments
are based on the relative flatness of the damping constants $\gamma (\omega
) $ as a function of frequency. This argument is closely related to the
notion of the flatness of the spectral density $G(\omega )$ in the SCL,
since the damping constants are proportional to $G(\omega )$ [see Eq.~(\ref%
{Dav})]. For example, below Eq. (2.95) in \cite{Carmichael:book} the author
argues that one can write down a Rabi frequency-independent MME for
resonance fluorescence since $\gamma (\omega _{A})$ and $\gamma (\omega
_{A}\pm \Omega _{\mathrm{R}})$ (where $\omega _{A}$ is the Bohr frequency)
differ by less than 0.01$\%$ at optical frequencies and reasonable laser
intensities. However, this ignores the corrections due to the Rabi frequency
to the operators $S_{k}(q,\omega )$ [Eq.~(\ref{18})]. This disagreement can
be traced to the question of at which point in the derivation it is safe to
neglect $\Omega _{\mathrm{R}}$; in \cite{Carmichael:book} this is done on
the basis of the flatness of $\gamma (\omega )$ before \textquotedblleft a
lot of tedious algebra\textquotedblright\ \cite{Carmichael:book}[p.48], but
our Floquet analysis shows that, in fact, one cannot neglect the Rabi
frequency relative to the Bohr frequency. This is relevant for our general
discussion since the \textquotedblleft \emph{finitely localized}
MME\textquotedblright\ which is the outcome of the Floquet analysis (see
next comment) actually exhibits a weak non-Markovian character. Such
deviations are, of course, important for FT-QEC, even if the effects are
small. We revisit this point in Section \ref{objections} below.

\noindent -- \emph{Are there any non-Markovian effects at work here?} It seems that
one should accept the \emph{generalized notion} of a quantum Markovian
master equation as the one given by Eqs.~(\ref{10}), (\ref{11}) and (\ref{2}%
), i.e., a master equation with a possibly time-dependent Lindblad
generator. In Davies' generalization to the time-dependent case \cite%
{Davies:78} (\textquotedblleft adiabatic WCL\textquotedblright ) the
dissipative generator $\mathcal{L}(t)$ depends on the Hamiltonian at the 
\emph{same time} $t$. This is a type of \textquotedblleft \emph{local}
generalized MME\textquotedblright . On the other hand, in the periodic WCL
treated here, the dissipative generator $\mathcal{L}(t)$ depends on the
Hamiltonians $H_{S}(u)$ from an interval, say $[0,t]$ ($t<\Theta $), as can
be seen from Eq.~(\ref{11}), which involves $\mathcal{U}(t,0)$. This is
therefore a type of \textquotedblleft \emph{finitely localized}
MME\textquotedblright , though one could argue that it exhibits a weakly
non-Markovian character because of this dependence of the dissipative
generator on the past. On the other hand, a non-Markovian master equation
(in the convolutionless formalism \cite{Breuer:book}) is also given by Eq.~(%
\ref{10}), but the generator is \emph{not} of Lindblad form [in particular,
it is not of the form (\ref{19})], and may depend on the Hamiltonian in the 
\emph{distant} past. The weight of distant past contributions depends on the
decay properties of $F(t)$ which are, generically, not exponential but
rather powerlike. In the WCL the non-Lindbladian terms vanish due to the
oscillating character of the $e^{i(\omega -\omega ^{\prime })u}$ terms in
Eq.~(\ref{eq:K2}).

\noindent -- \emph{The original Davies derivation}:\ We note that the Davies result
is a limit theorem which states that for a sufficiently small coupling
constant the WCL semigroup is a good approximation to the real dynamics.
However, Davies' theorem itself does not provide the conditions under which
a given physical coupling is \textquotedblleft small
enough\textquotedblright . In particular, one cannot extract from Davies'
theorem under what conditions the fast oscillating terms vanish. This can,
however, be done by a more heuristic analysis, as done above.

\subsubsection{WCL for an Arbitrary Pulse}

We now consider the case 
\begin{equation}
H_{C}(t)=H_{0}+f(t)H_{1},
\end{equation}%
i.e., an arbitrary driving field. This is, of course, the case of most
interest in FT-QEC. It follows from Fourier analysis that this case can be
treated qualitatively as a \textquotedblleft
superposition\textquotedblright\ of periodic perturbations discussed above.
For a single frequency $\Omega $ the validity of the Markovian approximation
is restricted by the condition (\ref{MA-per}):\ $t\gg \max \{1/(\omega
-\omega ^{\prime }+m\Omega )\}$. The discreteness of the frequencies $%
\{\omega \}$ and $\{m\Omega \}$ is key:\ it allows for condition (\ref%
{MA-per}) to be satisfied with finite $t$. A pulse $f(t)$ has a continuous
band of frequencies of width $\Gamma \simeq 1/\tau _{g}$ (where $\tau _{g}$
is the gate duration), with amplitudes (Fourier transform) $\hat{f}(\Omega )$%
, which add to and smear the effective Bohr spectrum $\{\omega \}$. If the
pulse is long (a slow gate) then only a narrow band appears, and the
smearing effect is unimportant. More precisely, if $1/\tau _{g}$ is much
smaller than the typical difference of the Bohr frequencies, the
\textquotedblleft energy quanta\textquotedblright\ $m\Omega $ [with $|m|$
restricted by the (typically small) dimension of the system Hilbert space]
cannot fill the gap between $\omega $ and $\omega ^{\prime }$ and the
condition (\ref{MA-per}) can be satisfied. This is our adiabatic
approximation. For fast pulses, when $1/\tau _{g}$ is comparable to $|\omega
-\omega ^{\prime }|$, the condition (\ref{MA-per}) cannot be fulfilled: the
effective Bohr spectrum becomes quasi-continuous and the denominator in
condition (\ref{MA-per}) becomes abitrarily small. The result is that the
WCL\ analysis breaks down and non-Markovian effects dominate.

Thus, the condition for the adiabatic limit (Markov approximation valid) is:
\textquotedblleft the width of the band is much smaller than the minimal
difference of the effective Bohr frequencies\textquotedblright . This is in
contradiction with the fast gate assumption, \textbf{A1}.

\subsection{Section Summary}

The main advantage of the MME (\ref{Dav}) is its consistency with
thermodynamics. Namely, as a consequence of the KMS condition (\ref{KMS})
and the condition (\ref{Dav1}), for a generic 
initial state the system tends
to its thermal equilibrium (Gibbs) state at the temperature of the heat
bath \cite{Alicki:87}.
(An important exception to this rule are states within a
decoherence-free subspace \cite{Zanardi:97c,LidarWhaley:03}, 
but these states are not generic due
to required symmetry properties of the
system-bath interaction.) Therefore the dissipative part of the generator
must depend strongly on the Hamiltonian dynamics. This is consistent with
the notion of a coarse-grained description familiar from the study of MMEs:
the bath needs a time much longer than $\max_{\omega _{kl}}1/\omega _{kl}$
to \textquotedblleft learn\textquotedblright\ the system's Hamiltonian in
order to drive it to a proper Gibbs state. In other words, \emph{the Markov
approximation is, equivalently, a long-time limit} (compared to $%
\max_{\omega _{kl}}1/\omega _{kl}$ -- the system's Bohr frequencies), and
one cannot expect this approximation to be valid at short times. However,
FT-QEC assumes operations on a time-scale that is short on the scale set by $%
\max_{\omega _{kl}}1/\omega _{kl}$.

Strictly speaking the MME (\ref{Dav}) is valid only when $H_{S}$ is not time
dependent. As we have shown, one can relax this by assuming slowly varying $%
H_{S}$, giving rise to an \textquotedblleft adiabatic MME\textquotedblright
, Eqs.~(\ref{10}), (\ref{11}) and (\ref{2}). However, to accept Eqs.~(\ref%
{10}), (\ref{11}) and (\ref{2}) as a genuine Markovian description is
somewhat of a stretch, since the real question is not whether one obtains
the Lindblad form, but rather \emph{how }$\mathcal{L}(t)$\emph{\ depends on
the Hamiltonians }$H_{S}(u)$\emph{, locally (i.e. }$u\simeq t$\emph{) or
nonlocally. For fast gates and generic environments the dependence is
non-local, involving memory effects. }In any case, the crucial condition
that must be satisfied for a (generalized) MME is Eq.~(\ref{MA-per}), which
implies that the average Bohr spectrum must be discrete. In essence, as long
as the applied control does not spoil this discreteness a (generalized) MME\
can be derived. On the other hand, \emph{this means that fast gates are
incompatible with the MME}, in violation of \textbf{A1} of FT-QEC theory.
The corollary: \emph{finite speed of gates implies non-Markovian effects}.

\section{Are the Standard FT-QEC Assumptions Internally Consistent?}

\label{consistency}

We now briefly summarize our examination of the assumptions of FT-QEC, in
light of the considerations above, and highlight where there may be internal
inconsistencies in FT-QEC. As discussed above, there are essentially two
rigorous approaches to the derivation of the MME: (i) the SCL, which is
compatible with arbitrarily fast Hamiltonian manipulations, but requires the
high-$T$ limit; (ii) the WCL, which is compatible with thermodynamics at
arbitrary $T$, but requires adiabatic Hamiltonian manipulations.

The standard theory of FT-QEC (excluding Refs.~\cite%
{Terhal:04,Aliferis:05,Aharonov:05}) requires a quantum computer (QC)
undergoing Markovian dynamics, supplemented with a constant supply of cold
and fresh ancillas. These assumptions are contradictory under the SCL, since
the QC would have to be at high-$T$, while the ancillas require low-$T$ on
the same energy scale $E$ (set by the Bohr energies of the system = computer
+ ancillas). Specifically, if we were to assume that for the ancillas too $%
kT\gg E$, they would quickly become highly mixed. If we insist that $kT\ll E$
for the ancillas, then by coupling them to the QC we can no longer assume,
in the SCL, that the total system = QC + ancillas is described by Markovian
dynamics.

If, on the other hand, we approach the problem from the (physically more
consistent) WCL, then \textbf{A3} is incompatible with \textbf{A1} (the
assumption of fast gates). Namely, in the WCL only adiabatic Hamiltonian
manipulations are allowed. Specifically, the Markov approximation in the WCL
requires a \emph{discrete} system (effective) Bohr frequency spectrum, such
that the condition$\ \tau _{g}\gg \max_{\omega _{kl}}1/\omega _{kl}$ can be
satisfied, hence violating the $\tau _{g}\omega _{B}=O(\pi )$ condition of 
\textbf{A1}. These conclusions are unavoidable if one accepts
thermodynamics, since they follow from seeking a Markovian master equation
that satisfies the KMS condition -- a necessary condition for return to
thermodynamic equilibrium in the absence of external driving. We take here
the reasonable position that a fault tolerant QC cannot be in violation of
thermodynamics.

\section{Possible objections to the inconsistency}

\label{objections}

In this section we analyze a list of possible objections to the
inconsistency we have pointed out.

\subsection{Is thermodynamics relevant?}

With respect to the SCL: \textquotedblleft \emph{Thermodynamics is
irrelevant (since a QC need not ever be in thermal equilibrium)}%
.\textquotedblright\ 

Note that we never claim that the QC is in thermal equilibrium; only the
bath is. This assumption is a simplification which allows us to use a single
parameter $T$ and therefore a single \textquotedblleft thermal memory
time\textquotedblright\ $\hbar /kT$. There is no reason to use a nonthermal
bath or many heat baths with different temperatures:\ this does not make the
spectral density flat and can only introduce more parameters.

\subsection{Doesn't the interaction picture save the day?}

With respect to the WCL: \textquotedblleft \emph{Suppose we have the
following Hamiltonian in the Schrodinger picture: }$%
H=H_{S}+H_{C}(t)+H_{SR}+H_{R}$\emph{\ where }$||H_{S}||\gg ||H_{C}||=$\emph{%
control Hamiltonian }$\gg ||H_{SR}||$\emph{. Then in the interaction picture
with respect to} $H_{S}$\emph{\ the term }$H_{C}$\emph{\ is dominant and hence can
implement fast gates. However, in the original Schr\"{o}dinger picture }$%
H_{C}$ \emph{is small and hence the adiabatic limit for the derivation of
the MME is satisfied. Thus we have an example where we can have fast gates
(in the interaction picture) and still the WCL can be satisfied so that the
Markovian limit can be reached. Moreover, this is the relevant limit relevant
for quantum optics, e.g., trapped ions.}\textquotedblright

There are a number of problems with this argument. First, one should be more
careful about the formulation of the condition for adiabaticity. It can be
stated as $|d\omega (t)/dt|\ll \omega (t)^{2}$, where $\omega (t)$ is a
\textquotedblleft relevant\textquotedblright\ Bohr frequency. Merely
comparing norms as above does not guarantee adiabaticity. Second, in the
quantum optics context we note the following. For three-level trapped ions
we have two Bohr frequencies: a large, time-independent $\omega _{1}$, and a
small, time-dependent $\omega _{2}(t)$ (degenerate levels splitting). Only $%
\omega _{2}$ is \textquotedblleft relevant\textquotedblright\ because it is
related to gates, and then the adiabatic condition implies that $|d\omega
_{2}(t)/dt|$ is correspondingly small, which contradicts the fast gate
condition \textbf{A1}. Third, the inequality $||H_{C}||\gg ||H_{SR}||$ is in
fact not satisfied in the Markovian WCL, where $||H_{SR}||$ diverges (one
should not confuse the small system-reservoir coupling parameter involved in
the van-Hove limit with the operator norm, which can be infinite).

\subsection{Doesn't quantum optics provide a counterexample?}

With respect to the WCL: \textquotedblleft \emph{Trapped ions and other
quantum optics systems provide a counter-example: a system experimentally
satisfying Markovian dynamics and allowing fast Rabi operations}%
.\textquotedblright\ 

We have already addressed quantum optical systems in Section \ref{Floquet}.
Let us add a few comments. We do not know of any quantum optics experiment
testing the Markov approximation with the accuracy relevant for FT-QEC (for
quantum dots, on the other hand, non-Markovian effects are very visible). We
know that for constant, and also for strictly periodic Hamiltonians (which
corresponds in quantum optics to a constant external laser field), the
Davies derivation can be applied (or extended, as in Section \ref{WCL}) and
the Markov approximation is applicable. The problem appears for fast gates.
It would be difficult to test the Markov approximation in this case with the
required accuracy, because, e.g., the results depend on the shape of the
pulse. A relevant example is resonance fluorescence, as described in \cite%
{Carmichael:book}[pp.43-61], and as discussed in Section \ref{Floquet}. The
damping effects are only present in the widths of spectral lines -- see \cite%
{Carmichael:book}[p.61, Fig. 2.5]. The Markov approximation gives
Lorentzians while non-Markovian dynamics may give rise to more complicated
lineshapes. Consider a 2-level atom like in \cite{Carmichael:book} Section
2.3.2., and in particular the final formula Eq. (2.96), which describes
resonance fluorescence via a MME. The author claims that for typical
parameters in quantum optics the dissipative part does not depend on the
Rabi frequency $\Omega _{\mathrm{R}}$ [recall our discussion in Section \ref%
{Floquet}]. Hence, as the gates are entirely related to $\Omega _{\mathrm{R}}
$, it appears that either fast or slow gates are possible. The argument is
based on the small ratio $\Omega _{\mathrm{R}}/\omega _{A}<10^{10}/10^{15}$
(where $\omega _{A}$ is the Bohr frequency). This is fine for replacing the
spectral density at $\omega _{A}\pm \Omega _{\mathrm{R}}$ by the density at $%
\omega _{A}$, but the subsequent argument that we can replace [in Eq.
(2.94)] $\Omega _{\mathrm{R}}$ by $0$ is inaccurate. This would be correct
only if the decay time $\tau =1/\gamma $ is short enough such that $\Omega _{%
\mathrm{R}}\tau \ll 1$. However, as explained in Section \ref{Floquet}, in
this case the Davies type averaging makes no sense physically. In fact,
typically for radiation damping $\tau =10^{-8}$s, and then $\Omega _{\mathrm{%
R}}\tau <100$ only. Hence for a fixed $\Omega _{\mathrm{R}}$ we do in fact
not have a simple Lindblad generator (of the type (2.96) in
\cite{Carmichael:book}), but rather a more complicated generator with
Lindblad 
operators depending on the Rabi frequency, as in Eq.~(\ref{16}). Again, in
the derivation of a proper generator an averaging over terms of the form $%
\exp (-i\Omega _{\mathrm{R}}t)$ must be performed. Therefore the condition
for the adiabatic approximation involves the Rabi frequency $\Omega _{%
\mathrm{R}}$ and cannot be satisfied for fast gates. For experiments based
on \emph{spectral measurements} the difference between the two types of
generators we have just discussed is probably irrelevant for many reasons;
however, the quantum state of the atom at a given moment is sensitive to a
small change in the Lindblad operators, and this is important in a fault
tolerant implementation of quantum logic gates.

\subsection{Is \textbf{A1} truly an assumption of FT-QEC?}

\label{adiabatic-gates}

With respect to the WCL: ``\emph{Doesn't \textbf{A1} impose an unnecessary
constraint on FT-QEC, in that gates are not required to satisfy the
condition }$\tau _{g}\omega =O(\pi )$\emph{?}''

In other words, one might argue in favor of slow gates, where instead the
condition is $\tau _{g}\omega \gg O(\pi )$. Such gates are certainly
relevant in the context of the adiabatic quantum computing (AQC) paradigm 
\cite{Farhi:01}, holonomic QC \cite{ZanardiRasetti:99,ZanardiRasetti:2000},
or topological quantum computing (TQC) \cite%
{Kitaev:97,Freedman:01,DasSarma:05}. We comment in more detail on AQC, HQC,\
and TQC in Section \ref{alternatives}. The question of interest to us is
whether an adiabatic gate satisfying $\tau _{g}\omega \gg O(\pi )$ is
applicable to the standard FT-QEC paradigm we are considering here, and
which is very different from AQC, HQC,\ and TQC.

First, let us clarify that by gates we mean one and two-qubit unitaries
picked from well-known discrete and small sets of universal gates\ \cite%
{Nielsen:book}. An algorithm is constructed via a sequence of such gates,
and computational complexity is measured in terms of the minimal number of
required gates. Of course one can instead join all gates used in a given
algorithm into a single unitary and call this a gate, but then one runs into
the problem of finding a relevant (physical) Hamiltonian and quantifying
computational complexity. For a given gate there are infinitely many
Hamiltonian realizations. Among these are fast ones (optimal) which satisfy $%
\tau _{g}\omega =O(\pi )$ and slow ones (adiabatic) satisfying $\tau
_{g}\omega \gg O(\pi )$ (all inequalities here are in the sense of orders of
magnitude). For example, consider a $\pi $-rotation. The fast (optimal)
realization satisfies $\tau _{g}\omega =\pi $ (compatible with \textbf{A1}),
while the slow (adiabatic) one satisfies $\tau _{g}\omega =\pi +2\pi n$ with 
$n\gg 1$ (contradicts \textbf{A1}).

Now, one may ask whether a slow realization of gates can prevent the
inconsistency with the WCL. We argue, based on computational complexity
considerations, that the answer to this question is negative. To see this,
note first that non-Markovian errors are uncorrectable in standard FT-QEC.
Therefore such non-Markovian, uncorrectable errors accumulate during the
computation (by definition, they are not corrected by \textquotedblleft
Markovian FT-QEC\textquotedblright ), and in order to keep them under
control, the probability of such errors per gate, $p_{\mathrm{non-M}}$,
should scale as 
\begin{eqnarray}
  p_{\mathrm{non-M}} &\sim& O[1/(\text{volume of algorithm})]
  \nonumber \\
  &=& O[1/(\text{input
size})^{\alpha }],
\end{eqnarray}
where $\alpha $ is some fixed power. Now, it follows from our discussion in
Section \ref{WCL} that the more adiabatic the evolution, the smaller is the
probability of the non-Markovian errors per gate. Therefore, if one writes
the adiabaticity condition as $\tau _{g}\omega >M$, where $M\gg 1$ is the
\textquotedblleft adiabatic slowness parameter\textquotedblright , then the
probability of non-Markovian errors should satisfy 
\begin{equation}
p_{\mathrm{non-M}}\sim O(1/M^{\beta }),
\end{equation}%
where $\beta $ is another fixed power [$\omega $ (the Bohr or Rabi
frequency) is limited essentially by the choice of physical system].
Comparing the two expressions for $p_{\mathrm{non-M}}$, we see that $M$ must
grow with input size. This means that if one works with adiabatic gates in
order to keep the dynamics (approximately) Markovian, the result is that one
must slow the gates in proportion to the input size (to some power). This,
however, violates the threshold condition of FT-QEC, in which the input size
and gate times are independent parameters (see, e.g., Theorem 12 in \cite%
{Aharonov:99}).

\subsection{Measurements}

With respect to both the WCL and the SCL:\ \textquotedblleft \emph{Recent
results on fault-tolerant QC using measurements only (e.g., \cite%
{Nielsen:04,Raussendorf:05}) render all the claimed problems irrelevant}%
.\textquotedblright\ 

Indeed, we have so far discussed only the problems with quantum logic gates.
Moreover, measurements are an integral part of FT-QEC theory as well, in
particular to reset and disentangle ancillas before they are introduced into
an error-correction circuit. Therefore some remarks on the use of
measurements are in order.

In the most advanced FT-QEC scheme of \cite{Aharonov:99}, measurements are
performed at the end of the computation. However, this approach demands a
high resource overhead, which may make it impractical. Therefore, more
recent proposals (e.g., \cite{Knill:05,Steane:04}) rely on feedback
mechanisms employing the results of quantum measurements. Those
\textquotedblleft measurements in the middle of
computation\textquotedblright\ are treated for simplicity as certain
von-Neumann projective measurements (but with efficiency $\ll 1$) satisfying
a \emph{repeatability condition}. The latter implies that the subsequent
measurements reduce the measurement error exponentially as their number
increases. This assumption should be carefully scrutinized, within realistic
Hamiltonian models of quantum measurement treated as a dynamical process.
Here, again one can expect that the tacit assumption of statistical
independence of repeated measurements is in conflict with the non-Markovian
character of the dynamics of open quantum systems.

As all proposed measurement schemes are based on electromagnetic
interactions, it should be possible to construct a rather general
Hamiltonian framework and apply it to various particular implementations.
Indeed, this has been done, e.g., for a single-electron tunneling (SET)
transistor coupled capacitively to a Josephson junction qubit \cite%
{Shnirman:98}. Rather than assuming that the measurement apparatus is
coupled to the system whenever measurements must be performed -- an option
which is hard to achieve in mesoscopic systems -- Ref.~\cite{Shnirman:98}
makes the reasonable assumption that the measurement apparatus is always
coupled to the system, but is in a state of equilibrium when it is not
needed. A measurement is then performed by driving the measuring device out
of equilibrium, in a manner that dephases the qubit to be measured. Generic
features emerging from this analysis are the existence of three different
time-scales characterizing the measurement: the dephasing time, the
measurement time (which may be longer than the dephasing time), and the
mixing time (the time after which all the information about the initial
quantum state is lost due to the transitions induced by the measurement).
Ref.~\cite{Shnirman:98} thus arrives at a criterion for a ``good'' quantum
measurement: the mixing time should be longer than the measurement time. A
time-scale analysis of measurements in optical systems, accounting for
spontaneous emission, can be found, e.g., in Ref.~\cite{Teich:89}. A fully
consistent analysis of FT-QEC should account for the existence of such
time-scales in a dynamic description of the measurement process. In
particular, it is important to set appropriate bounds on these time-scales,
so that they may be taken into account in a threshold calculation (an
analysis based on a stochastic error model was reported in Ref.~\cite%
{Steane:03}).

\subsection{Degenerate Qubits}

With respect to the SCL: \textquotedblleft \emph{Degenerate qubits
automatically satisfy the high }$T$\emph{\ limit since their intrinsic
energy scale vanishes}.\textquotedblright\ 

Examples of degenerate qubits are common, e.g., in trapped ion quantum
computing implementations where a pair of degenerate hyperfine states can
serve as a qubit, with an auxiliary third level used to implement quantum
logic gates via Raman transitions \cite{Wineland:98}. The case of degenerate
qubits is somewhat more subtle to analyze within the context we have
explained above. Naively, in such a case the high-$T$ limit is indeed
automatically satisfied, since the system energy scale is zero. Therefore it
appears that one could claim that the SCL version of the Markov
approximation is attainable. However, upon closer examination this still
seems problematic. Indeed, the vanishing of an energy scale for degenerate
qubits holds, strictly speaking, only for fully adiabatic techniques, e.g.,
HQC \cite{ZanardiRasetti:99,ZanardiRasetti:2000}. Otherwise transformations
between logical states are achieved by resorting to effective Hamiltonians
which involve \emph{virtual} transitions. For instance, if $|0\rangle $ and $%
|1\rangle $ denote degenerate qubit levels (e.g., hyperfine levels of an
ion), one can introduce far-detuned (e.g., laser) couplings of $|0\rangle $
and $|1\rangle $ with a third auxiliary level. Second order perturbation
theory then yields the effective Hamiltonian $H_{\mathrm{eff}}=-(\Omega _{%
\mathrm{R}}^{2}/\Delta )|1\rangle \langle 0|+\mathrm{h.c.}$, where $\Omega _{%
\mathrm{R}}$ and $\Delta $ are the laser Rabi coupling and detuning,
respectively. Therefore we see that an effective, small but non-vanishing,
energy scale $E_{1}:=\Omega _{\mathrm{R}}^{2}/\Delta $ is introduced. (Note
that in order for perturbation theory to be valid one must have $\Omega _{%
\mathrm{R}}\ll \Delta $, which in turn implies $E_{1}\ll \Delta $.) Yet
another energy scale is provided by the spectral width $E_{2}$ of the laser
pulse shape $\Omega _{\mathrm{R}}(t)$; in order to suppress unwanted \emph{%
real} transitions, one must impose in addition that $E_{2}\ll \Delta $. At
any rate, the appearance of these new system-energy scales implies that once
again the SCL-type contradiction applies. On the other hand, we can make
both $E_{1}$ and $E_{2}$ small at the price of lengthening the gating time ($%
\tau _{g}\simeq \max \{1/E_{1},\,1/E_{2}\}$). This implies, once again, an
adiabatic limit and the applicability of the WCL. Therefore it appears that
as long as one restricts manipulations to adiabatic ones (thus contradicting 
\textbf{A1}), quantum computing with degenerate qubits is possible even in
the Markovian limit. We expand on this viewpoint below.

\subsection{Impure Ancillas}

With respect to the SCL:\ \textquotedblleft \emph{Do ancillas really need to
be pure?}\textquotedblright\ 

What precisely is the role of the ancillas in QEC? A popular answer is that
they serve as an \textquotedblleft entropy sink\textquotedblright\ for the
errors accumulated during the quantum computation. This entropy in the
system arises from the entanglement between system and bath, and the role of
the ancillas is to remove this entanglement. I.e., in a perfect quantum
error correction step the entanglement between system and bath is
transferred to the ancillas and bath. A natural objection to our SCL-based
inconsistency is to claim that, in fact, ancillas need not be pure, or could
perhaps even be highly mixed. However, this is not supported by the
(current) standard theory of FT-QEC. Consider, e.g., an error correction
circuit based on the Steane 7-qubit code. It takes as input ancillas
prepared in the $|\psi \rangle _{a}=(|0_{L}\rangle +|1_{L}\rangle )/\sqrt{2}$
state, where $|0_{L}\rangle $ and $|1_{L}\rangle $ are codewords. The
physical qubits which comprise such ancillas, are coupled bitwise via CNOT
gates to the physical qubits making up the encoded data qubits in the
circuit. If instead we input an ancilla in a mixed state, this is equivalent
to inputting a classical mixture with erred codewords, e.g., $(1-p)|\psi
\rangle _{a}\langle \psi |+p|\phi \rangle _{a}\langle \phi |$, where $|\phi
\rangle _{a}$ is an erred codeword. If one of these errors is a phase-flip,
it feeds back (via the CNOT gates) into the data qubits, producing an error 
\cite{Gottesman:97a}. Without fault-tolerance this means that there are now
two errors (in the ancillas block and the data block), which may be more
than the code can handle. In FT-QEC theory such errors are accounted for,
but their magnitude is bounded from above (e.g., $p$ in the above example
must be small). We note that an ancilla which is initially entangled with
the data qubits (violating the assumption of being introduced into the
circuit in a tensor-product state) is essentially equivalent to the case of
an impure ancilla just described (tracing over the data qubits yields an
impure ancilla state).

A more general approach showing the importance of the assumption of pure
ancillas is the following (fairly standard account of QEC).

\noindent i) \textit{Preparation}.--

Let the initial state of system + reservoir + ancillas, with respective
Hilbert spaces $\mathcal{H}_{S},\mathcal{H}_{R},\mathcal{H}_{A}$, be: $\rho
_{SRA}^{0}=|\psi _{S}\rangle \langle \psi _{S}|\otimes |0_{R}\rangle \langle
0_{R}|\otimes \rho _{A}$, where we have allowed for ancillas in a mixed
state $\rho _{A}$.

\noindent ii) \textit{System-reservoir interaction (decoherence)}.-- 
\begin{equation}
\rho _{SRA}^{0}\overset{U_{SR}}{\longrightarrow }\rho
_{SRA}^{1}=\sum_{e,e^{\prime }\in \mathcal{E}}U_{e}|\psi _{S}\rangle \langle
\psi _{S}|U_{e}^{\dag }\otimes |e_{R}\rangle \langle e_{R}^{\prime }|\otimes
\rho _{A},
\end{equation}%
where $e$'s denote the \emph{errors} belonging to the set $\mathcal{E}$ that
the code $\mathcal{C}$ can correct, and where $|e_{R}\rangle $ are the
corresponding states of the reservoir. The error operators $U_{e}$ are
assumed to be unitary and with linear span of dimension $|\mathcal{E}|$.

\noindent iii) \textit{System-ancilla interaction (syndrome extraction)}.--

This interaction takes the form $U_{SA}=\sum_{e\in \mathcal{E}}\Pi
_{e}\otimes T_{e}$ where the $T_{e}$'s are unitaries over $\mathcal{H}_{A}$
such that $T_{e}|0_{A}\rangle =|e_{A}\rangle $ and $\Pi _{e}\cong I_{%
\mathcal{C}}\otimes |e\rangle \langle e|$ \footnote{%
We know that $\mathcal{H}_{S}\cong \mathcal{C}\otimes \mathcal{S}\oplus 
\mathcal{D}$ [$\mathcal{S}$=syndrome subsystem, dim$\mathcal{D}=|\mathcal{E}|
$; $\mathcal{D}$=remainder (=$0$ for subspace-based codes)] \cite%
{Knill:97b,Knill:99a,Zanardi:99d}.}: 
\begin{eqnarray}
\rho _{SRA}^{1} &\overset{U_{SA}}{\longrightarrow }&\rho _{SRA}^{2}  \notag
\\
&=&\sum_{e,e^{\prime }\in \mathcal{E}}U_{e}|\psi _{S}\rangle \langle \psi
_{S}|U_{e^{\prime }}^{\dagger }\otimes |e_{R}\rangle \langle e_{R}^{\prime
}|\otimes T_{e}\rho _{A}T_{e^{\prime }}.  \notag \\
&&
\end{eqnarray}

\noindent iv) \textit{Error recovery}.---

Unitary recovery is implemented via $\tilde{U}_{SA}=|\mathcal{E}%
|^{-1/2}\sum_{e\in \mathcal{E}}U_{e}^{\dagger }\otimes I_{R}\otimes
|e_{A}\rangle \langle e_{A}|$, where for unitarity we need $\langle
e_{A}|e_{A}^{\prime }\rangle =\delta _{e,e^{\prime }}$. By applying $\tilde{U%
}_{SA}$ and tracing over both $R$ and $A$ (assuming the $|e_{R}\rangle $'s
too are orthonormal) one obtains 
\begin{equation}
\rho _{S}^{\mathrm{out}}=\frac{1}{|\mathcal{E}|}\sum_{e,f\in {E}%
}U_{f}^{\dagger }U_{e}|\psi _{S}\rangle \langle \psi _{S}|U_{e}^{\dagger
}U_{f}\,\langle f_{A}|T_{e}\rho _{A}T_{e}^{\dagger }|f_{A}\rangle .
\end{equation}%
In the case of a pure ancillas $\rho _{A}=|0_{A}\rangle \langle 0_{A}|$ one
has $\langle f_{A}|T_{e}\rho _{A}T_{e}^{\dagger }|f_{A}\rangle =|\langle
f_{A}|e_{A}\rangle |^{2}=\delta _{f,e}$ and therefore the ideal case $\rho
_{A}^{\mathrm{out}}=|\psi _{S}\rangle \langle \psi _{S}|$ is recovered. One
can also consider the fidelity 
\begin{eqnarray}
F &:=&\langle \psi _{S}|\rho _{S}^{\mathrm{out}}|\psi _{S}\rangle   \notag \\
&=&|\mathcal{E}|^{-1}\sum_{e,f\in {E}}|\langle \psi _{S}|U_{f}^{\dagger
}U_{e}|\psi _{S}\rangle |^{2}\,\langle f_{A}|T_{e}\rho _{A}T_{e}^{\dagger
}|f_{A}\rangle .  \notag \\
&&
\end{eqnarray}%
Provided the error operators $U_{f}$ satisfy the condition for a
non-degenerate code $\langle \psi _{S}|U_{f}^{\dagger }U_{e}|\psi
_{S}\rangle =\delta _{f,e}$ \cite{Knill:97b}, one obtains $F=|\mathcal{E}%
|^{-1}\sum_{e\in \mathcal{E}}\langle e_{A}|T_{e}\rho _{A}T_{e}^{\dagger
}|e_{A}\rangle =\langle 0_{A}|\rho _{A}|0_{A}\rangle .$ Clearly, $F=1$ iff $\rho _{A}=|0_{A}\rangle \langle 0_{A}|,$ i.e., \emph{%
the ancillas are pure}. One can also consider non-unitary recovery via
ancilla measurements and conditional unitaries, with Kraus operators given
by $A_{e}=|\mathcal{E}|^{-1/2}U_{e}^{\dagger }\otimes I_{R}\otimes
|e_{A}\rangle \langle e_{A}|$. The conclusion that the ancillas' state must
be pure is unchanged.

We note that FT is obtained by adding concatenation and, in steps iii) and
iv), preparing and coupling encoded ancillas with the system in a suitable
way, e.g., as in the Steane-code example above. In this case it is
permissible to allow slightly impure ancillas, and relax the assumptions
that, in step ii) the environment couples only to the system, and in steps
iii,iv), the environment does not act. This formulation, however, does not
allow arbitrarily mixed-state ancillas, as argued in the Steane-code
example. While such a formulation of FT-QEC theory might still emerge (for
example, by using algorithmic cooling techniques \cite%
{Schulman:98,Schulman:05}, which, however, at present assume perfect gates),
it does not appear possible at present to relax the assumption of cold
ancillas.

\subsection{Hot QC, cold ancillas, and fast QC-ancilla interactions in the
SCL}

\label{Strong}

With respect to the SCL: \emph{\textquotedblleft One can keep the ancillas
coupled to a separate cold bath and then couple them for only a short time
to the QC: what matters then is the }$\emph{T}_{1}$\emph{\ timescale and
that one can be very long compared to the required ancilla-QC coupling
time\textquotedblright }.

Let us paraphrase this objection. If one can make $H_{SA}$ (system-ancillas)
very large then one could beat the rate of ancilla heating by strongly
coupling the QC and ancillas. I.e., suppose one would like to bring the
ancillas in from their cold reservoir to couple to the system, which is
coupled to a hot reservoir as required for the SCL. The ancillas then heat
up fast, but there is a timescale associated with this heating
(\textquotedblleft $T_{1}$\textquotedblright ), which one wishes to beat.
Now if one could make the system-ancilla coupling very strong then one
could, presumably, use the ancillas (e.g. for syndrome extraction) faster
than their heating rate, while they are still sufficiently pure for fault
tolerance purposes.

The simplest argument against this objection is the following. In the
setting of the objection, the QC is described by the SCL (high $T$) while
the ancillas are described by the WCL (low $T$). Strong and fast coupling
between the QC and the ancillas is unacceptable according to the WCL because
it is fast
(only adiabatic manipulations are allowed),
and according to the  SCL because it is strong
(``strong'' refers to the system's Hamiltonian part, while in the SCL this Hamiltonian is weak
in comparison with the system-bath coupling).

However, one could go on to argue that the ancillas are a different species
than the QC qubits, and in particular have a different intrinsic
(less
dense) energy scale, so that they are at low $T$ on the scale set by the QC\
qubits. In this case both ancillas and QC are described by the SCL. Then the
problem with the objection is the following: recall that in the SCL (see
Section \ref{SCL}) one must rescale $H_{SR}$ and $H_{AR}$ as $%
H_{SR}/\epsilon $ and $H_{AR}/\epsilon $ respectively, where here $R$
denotes the common reservoir the system and the ancillas are coupled to.
The heating rate is proportional to the square of the coupling
strength to the reservoir, i.e., to $1/\epsilon ^{2}$, and hence diverges in
the SCL. Therefore to beat the ancilla heating process via fast manipulation
of the system-ancilla coupling one would have to rescale $H_{SA}$ at least
by $1/\epsilon ^{2}$, but this contradicts the SCL derivation, where in fact
one must keep $H_{SA}$ fixed while rescaling $H_{SR}$. The reason for this
is that, in the SCL derivation, it is the system (now including the
ancillas) that sets the timescale against which reservoir correlations must
be accelerated.\footnote{%
Let us also consider the issue from the perspective of thermodynamics. This
is not really necessary, since the arguments above about the SCL are
rigorous, but is interesting in its own right. 
First, we remark that error correction should really be made to work at the
common lower (initial ancillas') temperature. Heating a part of a QC only to
be closer to the Markovian limit is a suboptimal strategy, because it increases
the strength of the noise and stimulates entropy production. 
Second, in standard FT-QEC heat (entropy) flows from the QC to the ancillas
only, while in reality one should expect a flow in both directions and
additionally an entropy production. To see this let us ignore for the moment
the coupling of the QC to the bath, and consider ancillas coupled to a heat
bath at temperature $T$. The ancillas can be kept pure by maintaining an
energy gap $\gg $ $kT$. Assume that the initial state of QC ($C$) and
ancillas ($A$) is a product state $|\psi _{C}\rangle \otimes |\psi
_{A}\rangle $. Switching on the interaction $H_{CA}$ we induce an
equilibration process (because the dynamics is Markovian) of $C+A$ towards
the Gibbs state $\rho _{CA}=\exp (-H_{CA}/kT)/Z$, which is \emph{entangled}
(here for simplicity $H_{CA}$ contains not only the interaction but is the
total Hamiltonian of $C+A$). After a single step of error correction the
total state of $C+A$ can be modeled by $(1-p)U|\psi _{C}\rangle \otimes
|\psi _{A}\rangle U^{\dag }+p\rho _{CA}$, where $U$ is unitary and $0<p\ll 1$%
. Then we switch off the interaction with the ancillas. Whatever we do next
separately with $C$ and $A$, we cannot eliminate the error due to the
entanglement present in the term $p\rho _{CA}$. This type of incorrectable
error accumulates and destroys FT-QEC. This is the back flow of entropy from
the ancillas bath to the QC, mentioned above.}

\section{Alternatives to Markovian FT-QEC}

\label{alternatives}

\subsection{Nature of the non-Markovian errors in the WCL}

While we have pointed out that, in the WCL, the application of fast gates is
likely to violate the conditions required for Markovian dynamics to persist,
we have not been specific about the type of non-Markovian effects that will
emerge. It is well known that FT-QEC is capable of dealing with errors that
change due to the application of gates. Namely, assume (slightly) faulty
gates correcting a specific error model described by a CP map $\Lambda$
[recall Eq.~(\ref{eq:Lambda})], are applied in sequence, $\Lambda
U_{N}^{\prime }\Lambda U_{N-1}^{\prime }\cdots \Lambda U_{1}^{\prime }$, and
these gates are (in some appropriate norm) close to the ideal gates $%
\{U_{i}\}_{i=1}^{N}$, i.e., for all $i$, $||U_{i}^{\prime }U_{i}^{\dag
}-I||\ll 1$. Then by inserting $U_{i}^{\dag }U_{i}$'s everywhere one obtains
the new sequence $\Lambda_{N}U_{N}\Lambda_{N-1}U_{N-1}\cdots
\Lambda_{1}U_{1} $, where $\Lambda_{i}:=\Lambda U_{i}^{\prime }U_{i}^{\dag }$%
, and FT-QEC is capable of dealing with such a (gate-modified) error model.
However, the non-Markovian effects that arise due to the application of fast
gates in the WCL, will in general \emph{not} be describable by a simple
time-local modification such as $\Lambda\rightarrow \Lambda U_{i}^{\prime
}U_{i}^{\dag } $. This can be worked out, e.g., using the methods of Ref.~%
\cite{Breuer:02}.

In order to formulate consistent alternatives to standard, Markovian FT-QEC
theory, it seems useful to start with a Hamiltonian formulation. As the
discussion below will illustrate, it appears that a hybrid approach will be
necessary, which combines alternatives to standard QC with a new version of
FT-QEC.

\subsection{Adiabatic Quantum Computing (AQC)}

We keep \textbf{A2} and \textbf{A3}, discard \textbf{A1} (fast gates), and
work in a purely adiabatic mode, thus permitting a consistent WCL. This may
indeed be possible using the adiabatic quantum computing (AQC) approach of
Farhi et al. \cite{Farhi:01}. At present there is little understanding of
the fault-tolerance of AQC. Some recent works explore AQC in the presence of
decoherence and/or control errors \cite%
{Childs:02,Shenvi:03,Roland:04,Aberg:04,SarandyLidar:05,aberg-2005-72}.
Indeed, the subject of the adiabatic approximation in open quantum systems
has only very recently been addressed \cite{SarandyLidar:04}, and used to
study AQC in open systems \cite{SarandyLidar:05}. Error correcting codes for
AQC\ were introduced very recently in \cite{Jordan:05}, but the
corresponding universal Hamiltonians involve many-body interactions (four
and six-body for 1-local and 2-local errors, respectively).

\subsection{Holonomic Quantum Computing (HQC)}

Another possibility for keeping \textbf{A2} and \textbf{A3}, and discarding 
\textbf{A1}, is provided by HQC \cite{ZanardiRasetti:99,ZanardiRasetti:2000}%
. HQC is an adiabatic scheme which relies on Abelian or non-Abelian
geometric phases to implement quantum logic gates. Quantum information is
encoded in a \emph{degenerate} set of eigenstates of a Hamiltonian depending
on a set of controllable parameters, e.g., external laser fields (recall our
discussion of degenerate states above). When these are adiabatically driven
along a suitable closed path, the initial quantum state is transformed by a
non-trivial unitary transformation (holonomy) that is geometrical in nature.
The key point is that the geometrical nature of the quantum holonomies is
believed to render HQC inherently robust against certain kinds of errors.
This alleged fault-tolerance has only recently been seriously begun to be
examined \cite{Solinas:04}; the emerging picture is that, while stability
against decoherence must still be assessed, HQC seems to exhibit a strong
robustness against stochastic errors in the control process generating the
required adiabatic loops \cite{Zhu:04}. Moreover, in the adiabatic
limit of Markovian dynamics it has been show that the geometric phase of a
single qubit coupled to a magnetic field is robust against both
dephasing and spontaneous emission (but not against bit
flips) \cite{SarandyLidar:05a}. Nevertheless, since deviations from strict
adiabaticity are inevitable, and adiabaticity is particularly challenging to
satisfy in open quantum systems \cite{SarandyLidar:04}, it is tempting to
combine HQC with FT-QEC in order to address the performance of HQC in the
presence of decoherence errors. Alternatively, we note that a hybrid
approach that seems to be rather promising is the embedding of HQC within a
DFS \cite{WuZanardiLidar:05}. This amounts to realizing a set of universal
quantum gates, acting on a DFS, by means of non-abelian quantum holonomies.
This strategy brings together the \textquotedblleft best of two
worlds\textquotedblright : the quantum decoherence avoidance virtues of DFSs
and the fault-tolerance of the all-geometric holonomic control. It is
possible that such an approach can be implemented for quantum information
processing in, e.g., trapped ions and quantum dots.

\subsection{Topological Quantum Computing (TQC)}

A robust way of performing quantum computations is based on excitations with
fractional statistics, since they have several fault-tolerant properties
built in. This idea is known as topological quantum computing (TQC) \cite%
{Kitaev:97,Freedman:01,DasSarma:05}. Physical realizations of the simplest
versions of TQC have been considered in the literature, using, e.g.,
rotating Bose-Einstein condensates \cite{Paredes:01} and superconducting
circuits \cite{Ioffe:02}. Let $\mathcal{C}$ denote the manifold of quantum
codewords. Strikingly, in TQC, one can have a trivial Hamiltonian, e.g., $%
H|_{\mathcal{C}}=0$, but nevertheless obtain non-trivial quantum evolution
due to the existence of an underlying topological \emph{global} structure
(boundary conditions). Quantum encoding is typically performed in a properly
designed degenerate ground state $\mathcal{C}$. This fact implies, for low
enough temperature, an exponential suppression of errors on encoded quantum
information due to thermal fluctuations. More importantly, topological
features can render such a ground state stable against errors represented by 
\emph{local} operators, namely error operators that do not involve a number
of qubits of the order of the size of the system. For example, in the
so-called toric codes \cite{Kitaev:97,Kitaev:book}, qubits are encoded in
the ground-state manifold of a lattice of interacting spins in such a way
that degenerate ground states are mutually connected only via high powers
(scaling linearly with lattice size) of local operators. Thus, here the
fault-tolerance properties are already built-in at the \emph{physical level}%
. However, while one can argue that topological encoding provides a stable
and passive quantum memory, it is not self-correcting as in today's
\textquotedblleft effectively naturally fault-tolerant\textquotedblright\
classical architectures (see Ref.~\cite{Bacon:05} for an eloquent exposition
of this point). Moreover, it is important to realize that as far as we know,
in its present state TQC still requires active intervention, in the form of
FT-QEC, when one tries to compute fault-tolerantly. Indeed, Preskill writes
in Ref.~\cite{Preskill:TQC-notes}[p.62], \textquotedblleft It is therefore
implicit that the temperature is small enough compared to the energy gap of
the model that thermally excited anyons are too rare to cause trouble, that
the anyons are kept far enough apart from one another that uncontrolled
exchange of charge can be neglected, and in general that errors in the
topological quantum computation are unimportant. If the error rate is small
but not completely negligible, then the standard theory of quantum fault
tolerance can be invoked to boost the accuracy of the simulation as
needed\textquotedblright . Ref.~\cite{Dennis:02} takes this approach and
explicitly lists \textbf{A2} and \textbf{A3} as necessary requirements for
fault-tolerant TQC. In contrast, \textbf{A1} is definitely \emph{not}
required in TQC:\ one performs computations by adiabatically dragging
quasiparticles around one another, and these operations must be slow
relative to the gap between the ground state and the first excited state.
The larger the gap the easier it is to satisfy this adiabaticity condition,
so this requirements is compatible with the thermal suppression of errors
mentioned above. In addition, TQC requires time-dependent controls to read
out the encoded data (Ref.~\cite{Freedman:00a} shows that all measurements
can be postponed until the readout of the final result of the computation).
However, a fully Hamiltonian analysis of the fault-tolerance of such
measurements is still lacking. Nevertheless, one could argue that the error
rate in a topological quantum computer could be made arbitrarily small by
increasing the system size and careful engineering, so that (similarly to
today's self-correcting, fault-tolerant classical computers), one could
ultimately perform TQC without any active intervention other than read-out
of the encoded data. An interesting, recent development in this direction
was reported in Ref.~\cite{Bacon:05}, which suggests that certain
three-dimensional quantum spin-lattices might be self-correcting.

\subsection{Non-Markovian Quantum Computing}
\label{nonM-FTQEC}

We keep \textbf{A1} and \textbf{A2} but discard \textbf{A3} (the Markov
approximation) at least in part. This appears to be a reasonable approach in
many cases, since the Markov approximation is clearly a highly idealized
limit (though it does hold remarkably well in some optical systems and in
liquid state NMR). Indeed, the degree of accuracy to which the Markov
approximation must be satisfied has been quantified, e.g., by Steane in \cite%
{Steane:04}: the probability of an uncorrectable (i.e., non-Markovian) error
per gate must be $<$ $10^{-10}$ for a computation involving $10^{9}$ gates
(this probability must scale with the input size, as explained in Section %
\ref{adiabatic-gates}). Alternative approaches to dealing with non-Markovian
baths are therefore of interest. For example, the papers \cite%
{Terhal:04,Aliferis:05,Aharonov:05} present an extension of FT-QEC theory to a
non-Markovian setting. We offer in this context the following observations:

\noindent 1. An important ingredient carried over directly and without
change from Markovian FT-QEC theory, is the crucial role of the fresh and
nearly pure ancillas \cite{Terhal:04,Aliferis:05,Aharonov:05}. We believe that the detailed mechanism
for introducing and discarding ancillas at specific times should be
reconsidered within a fully Hamiltonian framework.

\noindent 2. As recognized and discussed in \cite{Terhal:04}, the important
assumption of a small norm of the system-bath interaction Hamiltonian (e.g.,
Eq.~(58) in Ref.~\cite{Aliferis:05}) is not satisfied for some standard
models of open systems. For example, a linear coupling to a bosonic heat
bath involves unbounded interaction Hamiltonians and a high-frequency
cutoff. In general, the assumption of a small norm of the system-bath
interaction Hamiltonian is much stricter than the WCL and is not satisfied
for most standard models of reservoirs.

Another approach to fault-tolerance in a non-Markovian setting is the
recently developed time-concatenated dynamical decoupling method \cite%
{KhodjastehLidar:04} (see also \cite{Viola:02} for a version of dynamical
decoupling with bounded-strength controls). However, comment 2. above about
the small norm of the system-bath interaction Hamiltonian applies here as
well. Therefore more general methods are required to deal with the full
scope of baths one can expect in quantum computing implementations. A
promising possibility in this direction is to incorporate fault-tolerant
dynamical decoupling in a feedback loop.

\section{Conclusions}

\label{conc}

We have listed a set of minimal assumptions made in the theory of
fault-tolerant quantum error correction (FT-QEC): 1) fast gates (on the
timescale set by the inverse of the relevant Bohr or Rabi frequency), 2) a
supply of fresh and nearly pure ancillas, 3) a Markovian bath.

We have also reviewed the only two known rigorous general limits leading to
Markovian dynamics: the singular coupling limit (SCL), which involves taking
a high temperature limit, and the weak coupling limit (WCL), which requires
either a constant or an adiabatic system Hamiltonian, and averaging over
long times in comparison with the inverse of the relevant Bohr frequency.
These two limits allow one to replace the reservoir autocorrelation function
by a Dirac delta, which leads to the Markovian limit.

A close examination of the assumptions of FT-QEC has led us to conclude that
assumption 3 can be sustained together with assumption 1 in the SCL, and
together with assumption 2 in the WCL. However, it is not possible to
maintain all three assumptions in either the SCL or the WCL. We therefore
conclude that, at present, there exists an inconsistency in the formulation
of the theory of FT-QEC for Markovian baths. We have also listed a number of
alternatives to Markovian FT-QEC which, from the point of view adopted here,
are free of inconsistencies. However, none of these alternatives is so
comprehensive as to include the full range of errors one might expect in a
full-scale implementation of quantum computing. In particular, recent
results on fault tolerance in non-Markovian settings \cite%
{Terhal:04,Aliferis:05,Aharonov:05,KhodjastehLidar:04}, while representing a
significant step forward, make a crucial assumption about the smallness of
the norm of the system-bath interaction Hamiltonian, which severely
restricts the class of physical reservoirs.

\acknowledgments We thank Dave Bacon, Andrew Doherty, Daniel Gottesman, Hideo Mabuchi, John
Preskill, Alireza Shabani, and Barbara Terhal for very useful discussions
(though this does not imply their agreement with our conclusions). Their
insightful comments helped us sharpen our critique and formulate the
questions in Section~\ref{objections}. 

R.A. thanks for the support from the Polish Ministry of Science and
Information Technology- grant PBZ-MIN-008/P03/2003 and the EC grant RESQ IST-2001-37559, D.A.L. thanks the Sloan
Foundation for a Research Fellowship and the DARPA-QuIST\ program for
support. P.Z. acknowledges support by the European Union FET project TOPQIP
(Contract No. IST-2001-39215).


\end{document}